\documentclass[prd,nofootinbib,twocolumn,showpacs,preprintnumbers,
amsmath,amssymb]{revtex4}

\usepackage{graphicx}
\usepackage{multirow,dsfont}
\usepackage{mathrsfs}  
\usepackage{slashed}  

\newcommand{\anti}[1]{{\overline{#1}}}
\newcommand{\ds}[1]{\slashed{#1}}
\newcommand{\rmt}{{\scriptscriptstyle\top}} 
\newcommand{\eq}[1]{\begin{equation}#1\end{equation}}
\newcommand{\eqs}[1]{\setlength\arraycolsep{2pt}\begin{eqnarray}#1\end{eqnarray}}
 
\newcommand{\deru}[2]{\frac{{\rm d} #1}{{\rm d} #2}}  
     
\newcommand{\re}[0]{{\rm Re}}

\newcommand{\diag}{{\rm diag}}

\newcommand{\id}[0]{\mathds{1}}
\newcommand{\refer}[1]{(\ref{#1})}
\newcommand{\nn}[0]{\nonumber}

\newcommand{\al}{\alpha}
\newcommand{\be}{\beta}
\newcommand{\de}{\delta}
\newcommand{\De}{\Delta}
\newcommand{\ga}{\gamma}
\newcommand{\Ga}{\Gamma}
\newcommand{\si}{\sigma}
\newcommand{\Si}{\Sigma}

\newcommand{\la}{\lambda}
\newcommand{\La}{\Lambda}

\newcommand{\ep}{\epsilon}
\newcommand{\vp}{\varphi}
\newcommand{\cphi}{\varphi}
\newcommand{\vep}{\varepsilon}

\newcommand{\ze}{\zeta}

\newcommand{\cC}[0]{{\mathcal{C}}}
\newcommand{\cD}[0]{{\mathcal{D}}}

\newcommand{\cF}[0]{{\mathcal{F}}}

\newcommand{\cK}[0]{{\mathcal{K}}}
\newcommand{\cL}[0]{{\mathcal{L}}}
\newcommand{\cM}[0]{{\mathcal{M}}}
\newcommand{\cN}[0]{{\mathcal{N}}}
\newcommand{\cO}[0]{{\mathcal{O}}}

\newcommand{\cY}[0]{{\mathcal{Y}}}

\newcommand{\sG}[0]{{\mathscr{G}}}

\newcommand{\sM}[0]{{\mathscr{M}}}

\newcommand{\sY}[0]{{\mathscr{Y}}}
\newcommand{\sZ}[0]{{\mathscr{Z}}}

\newcommand{\bM}[0]{{\mathbb{M}}}

\newcommand{\bV}[0]{{\mathbb{V}}}

\newcommand{\bX}[0]{{\mathbb{X}}}

\newcommand{\koment}[1]{}
\newcommand{\nothing}[1]{}
\newcommand{\kA}[0]{}
\newcommand{\kB}[0]{}
\newcommand{\kC}[0]{}
\newcommand{\kD}[0]{}
\newcommand{\kE}[0]{}
\newcommand{\kF}[0]{}
\newcommand{\kG}[0]{}

\newcommand{\kI}[0]{}
\newcommand{\kJ}[0]{}
\newcommand{\kK}[0]{}
\newcommand{\kL}[0]{}
\newcommand{\kM}[0]{}
\newcommand{\kN}[0]{}
\newcommand{\kO}[0]{}

\def\beq{\begin{equation}}
\def\eeq{\end{equation}}
\def\bea{\begin{eqnarray}}
\def\eea{\end{eqnarray}}
\def\ba{\begin{array}}
\def\ea{\end{array}}
\def\bit{\begin{itemize}}
\def\eit{\end{itemize}}
\def\pa{\partial}

\def\nnn{\nonumber\\}

\def\la{\lambda}
\def\La{\Lambda}

\def\re{{\rm e}}
\def\rd{{\rm d}}
\def\ri{{\rm i}}
\def\GeV{{\rm GeV}}
\def\TeV{{\rm TeV}}
\def\eV{{\rm eV}}
\def\MeV{{\rm MeV}}

\def\da{{\dot\alpha}}

\def\db{{\dot\beta}}

\def\dd{{\rm d}}

\def\vp{\varphi}

\def\vp{\varphi}

\def\rd{{\rm d}}
\def\ri{{\rm i}}

\def\cL{{\cal{L}}}

\def\cO{{\cal O}}
\def\cC{{\cal C}}
\def\cM{{\cal M}}
\def\cF{{\cal F}}
\def\cK{{\cal K}}

\def\tm{{\tt m}}

\def\MPL{M_{\rm PL}}
\def\RCI{{\rm R}_{\rm CI}}

\makeatletter
\def\underbracket{%
  \@ifnextchar [ %
    {\@underbracket}%
    {\@underbracket [\@bracketheight]}}
\def\@underbracket[#1]{%
  \@ifnextchar [ %
    {\@under@bracket[#1]}%
    {\@under@bracket[#1][0.4em]}}
\def\@under@bracket[#1][#2]#3{
  \mathop {%
    \vtop {%
      \m@th \ialign {%
        ##\crcr $\hfil \displaystyle {#3}\hfil $%
       \crcr \noalign %
       {\kern 3\p@ \nointerlineskip }%
        \upbracketfill {#1}{#2}
       \crcr \noalign %
       {\kern 3\p@ }%
     }%
   }%
  }%
  \limits%
}
\def\upbracketfill#1#2{%
  $\m@th \setbox \z@ \hbox {$\braceld$}
  \edef\@bracketheight{\the\ht\z@}\bracketend{#1}{#2}
  \leaders \vrule \@height #1 \@depth \z@ \hfill
  \leaders \vrule \@height #1 \@depth \z@ \hfill%
  \bracketend{#1}{#2}$%
}
\def\bracketend#1#2{\vrule height #2 width #1\relax}

\begin{document}
\title{Conformal Standard Model, Leptogenesis and Dark Matter}
\author{Adrian Lewandowski$^{1,}$\footnote{Present address:\\
\emph{Albert Einstein Center for Fundamental Physics,\\ 
Institute for Theoretical Physics, University of Bern, \\
Sidlerstrasse 5, CH-3012 Bern, Switzerland\\ }}, 
Krzysztof A. Meissner$^2$ and Hermann Nicolai$^1$}
\affiliation{
\phantom{aaa}\\
\phantom{aaa}\\
$^1$Max-Planck-Institut f\"ur Gravitationsphysik
(Albert-Einstein-Institut)\\
M\"uhlenberg 1, D-14476 Potsdam, Germany\\
\phantom{aaa}\\
$^2$Faculty of Physics, University of Warsaw\\
ul. Pasteura 5, 02-093 Warsaw, Poland\\
}

\begin{abstract} 
The Conformal Standard Model (CSM) is a minimal extension of the Standard Model 
of Particle Physics based on the assumed absence of large intermediate scales between
the TeV scale and the Planck scale, which incorporates only right-chiral neutrinos and a new
complex scalar in addition to the usual SM degrees of freedom, but no other features
such as supersymmetric partners. In this paper, we present a comprehensive 
quantitative analysis
of this model, and show that all outstanding issues of particle physics proper can in 
principle be solved `in one go' within this framework. This includes in particular the
stabilization of the electroweak scale, `minimal' leptogenesis and the explanation of 
Dark Matter, with a small mass and very weakly interacting Majoron as the Dark Matter
candidate (for which we propose to use the name `minoron').  The main testable prediction 
of the model is a new and almost sterile scalar boson that would manifest itself as 
a narrow resonance in the TeV region. We give a representative 
range of parameter values consistent with our assumptions and with observation.

\end{abstract}
\pacs{}

\maketitle

\vspace{0.2cm}
\section{Introduction}
The conspicuous absence of any hints of `new physics' at LHC, and, more pertinently,
of supersymmetric partners and exotics \cite{SUSY.png,Exotics.png} has prompted
a search for alternative scenarios beyond the Standard Model (SM)  
based on the hypothesis that the SM could survive essentially  {\em as is} 
all the way to the Planck scale, modulo `minor' modifications of the type discussed here,
see  \cite{Hemp,Shap,a1,KM,a2,CW5a,CW4,CW1,CWzz,CW2,SW,a5,noCWxx,noCWyy,U3model,nr1,nr2,nr3,nr4,nr5}
for a (very incomplete) list of
references. In this paper we follow up on a specific proposal along these lines which is 
based on our earlier work \cite{KM}, and demonstrate that this proposal in principle allows 
for a comprehensive treatment of all outstanding problems of particle physics proper.
This list includes perturbativity and stability of the model up to the Planck scale and an 
explanation of leptogenesis and the nature of Dark Matter\footnote{We thus exclude
 Dark Energy and the origin of inflation from this list; these are problems that, in our view, 
 will likely require a proper theory of quantum gravity for their complete resolution. 
 We note, however, that Higgs inflation \cite{Bezrukov:2008ut} can be easily incorporated
 into the present model \cite{Kw}.}, in a way which is in 
complete accord with the fact that LHC has so far seen nothing, and furthermore 
appears to be fully consistent as a relativistic QFT all the way up to the Planck scale
$\MPL$ (for which we use the reduced value  $\MPL \approx 2.4\cdot 10^{18}$ GeV). 
The consistency up to that scale, but not necessarily beyond, is in accord with
our essential assumption that, at the Planck scale, an as yet unknown UV complete theory of 
quantum gravity and quantum space-time takes over that transcends 
space-time based relativistic QFT. Importantly, 
the present approach is essentially `agnostic' about what this theory is.

Added motivation for the present investigation comes from very recent LHC results which
indicate that the low energy supersymmetry paradigm which has dominated much 
of particle physics over the past three decades is close to failure, unless one resorts to the
more exotic possibility that `low energy' ($N\!=\!1$) supersymmetry is broken at a very high scale.
In our opinion, however, the latter option would defeat the original purpose of solving the hierarchy 
problem, and thus lack the plausibility of the original Minimal Supersymmetric Standard Model (MSSM). One crucial question is therefore
how the `naturalness' of the electroweak scale can be explained without appealing
to supersymmetric cancellations. In this paper we offer one possible such alternative 
explanation based on \cite{CLMN}; another possibility which bears some resemblance 
to the present scheme as far as physics up to $\MPL$ is concerned (but not beyond) is 
to invoke asymptotic safety, see {\em e.g.} \cite{Reuter-review,SW,Litim}.

In its original form the model proposed in \cite{KM} tried to exploit the fact that, with 
the exception of the scalar mass term that triggers spontaneous breaking of  electroweak 
symmetry, the SM Lagrangian is classically conformally invariant. For this
it relied on the Coleman-Weinberg (CW) mechanism \cite{CW}  to break
electroweak symmetry and to argue that mass scales can be generated 
purely by the quantum mechanical breaking of classical conformal 
invariance. In this paper a modified version of this model is presented  which has explicit 
mass terms but which is still conformal in the sense that it postulates the absence 
of any intermediate scales between 1 TeV and the Planck scale $\MPL$
-- hence the name {\em Conformal Standard Model}.  The model nevertheless
achieves a stabilization of the electroweak hierarchy thanks to an alternative proposal 
for the cancellation of quadratic divergences presented in \cite{CLMN}, and it is in
this sense that we speak of {\em softly broken conformal symmetry} (SBCS). 
This term is meant to comprise three main assumptions, 
namely: (i) the avoidance of quadratic divergences, (ii) the smallness 
(w.r.t. the Planck scale) of all dimensionful quantities, and (iii) the smallness of all 
dimensionless couplings up to $\MPL$. With these assumptions 
the model is indeed `almost conformal', and the quantum 
mechanical breaking of conformal invariance (as embodied in the CW correction 
to the effective potential) remains a small correction over this whole range 
of energies. Here, we extend our previous considerations towards a more 
complete picture, in an attempt to arrive at a minimal comprehensive solution 
to the outstanding problems of particle physics, in a way that remains compatible 
with all available LHC results. More specifically, we here focus on the question whether
the model can offer a viable explanation for leptogenesis and the origin of Dark Matter. 
The main message of this paper, then, is that indeed these problems can 
be solved at least in principle within this minimal SBCS scheme. However, as we
said, no attempt will be made towards a solution of cosmological constant problem, 
nor inflation or Dark Energy, as these probably require quantum gravity.


The present paper thus puts together different ideas most of which have already appeared
in different forms in the literature  (in particular, conformal symmetry, extra sterile scalars and 
`Higgs portals', low mass heavy neutrinos, resonant leptogenesis, and quantum gravity 
induced violations of the Goldstone theorem), 
though, to the best of our knowledge, never in the combination proposed and elaborated 
here. Let us therefore summarize the distinguishing special features and assumptions 
underlying the present work:

\bit
\item There are no large intermediate scales between the TeV scale and the 
         Planck mass; in particular there is no grand unification nor GUT
         scale physics. 
\item There is no low energy supersymmetry; instead the electroweak hierarchy
         is stabilized by the alternative mechanism for the cancellation of
         quadratic divergences proposed in \cite{CLMN}.\footnote{However, this
         assumption by no means excludes the possibility that (extended) supersymmetry does play 
         an essential role {\em at the Planck scale} to ensure finiteness (UV completeness) 
         of a unified theory of quantum gravity.}
\item The consistency of the model up to the Planck scale is ensured by
         demanding absence of Landau poles and of instabilities or meta-stabilities
         up to that scale. Possible pathologies that might appear if the model 
         is extrapolated {\em beyond} that scale are assumed to be taken care of by quantum
         gravity, hence are not relevant for the present analysis.
\item The model naturally incorporates resonant leptogenesis \cite{PilafRL1,RL2,RL3,RL4} 
         with low mass heavy neutrinos, where we show that a range of parameters
         exists which meets all requirements. Furthermore, the predictions of the model do 
         not in any way affect the SM tests that have so far  confirmed the SM {\em as is}.
\item The Majoron, {\em i.e.} the Goldstone boson of spontaneously 
         broken lepton number symmetry, is assumed to acquire a small 
         mass $\sim 10^{-3}$ eV due to a (still conjectural) folklore theorem according 
         to which there cannot exist unbroken continuous global symmetries in quantum gravity,
         as a consequence of which it becomes a possible Dark Matter candidate (whose
         abundance comes out with the right order of magnitude subject to our assumptions).
         The ensuing violation of the Goldstone Theorem entails calculable couplings 
         to SM particles from radiative corrections, which are naturally very small.
\item The main testable prediction of the model is a new scalar resonance at $\cO(1\,\TeV)$ 
         or even below that is accompanied by a (in principle measurable) reduction of 
         the decay width of the SM-like Higgs boson. The couplings of the new 
         scalar to SM particles are strongly suppressed in comparison with those of the 
         SM Higgs boson by a factor $\sin\be$, where the angle $\be$
         parametrizes the mixing between the SM Higgs boson and the new scalar.
         The only new {\em fermionic} degrees of freedom are three right-chiral neutrinos. 
\item Because our model contains no new scalars that carry charges under SM gauge symmetries  
         it can be easily discriminated against many other models with an 
         enlarged scalar sector, such as two doublet models.
\eit

We note that a comprehensive `global' and quantitative analysis of the type performed here
would be rather more cumbersome, or even impossible, for more extensive
scenarios beyond the SM with more degrees of freedom and more free parameters. 
For instance, even with a very
restricted minimal set of new degrees of freedom and parameters as in the present 
setup, closer analysis shows that in order to arrive at the desired 
physical effects such as resonant leptogenesis with the right order of magnitude 
for the lepton asymmetry a very careful scan over parameter space is required, as the physical 
results can depend very sensitively on {\em all} parameters of the model, so
some degree of fine-tuning may be unavoidable.

The structure of this paper is as follows. In section II we describe the basic
properties of the model, and explain how to maintain perturbativity and stability
up to the Planck scale.  Section III is devoted to a detailed discussion of leptogenesis
in the CSM, and shows that a viable range of parameters exists for which resonant
leptogenesis can work. In section IV we discuss $(B-L)$ breaking and the possible role
and properties of the associated pseudo-Goldstone boson (`minoron') as a Dark Matter candidate. 
Although we present a representative range of parameters consistent with all our assumptions
and with observations, we should emphasize that our numerical estimates are still quite 
preliminary. Of course, these estimates could be much improved if the new scalar 
were actually found and its mass value measured.
For the reader's convenience we have included an appendix explaining 
basic properties of neutrino field operators
in Weyl spinor formalism.

\section{The CSM Model}
The Conformal Standard Model (CSM) is a minimal extension of the Standard Model 
that incorporates right-chiral neutrinos and an additional complex scalar field, which
is charged under SM lepton number, like the right-chiral neutrinos, and generates 
a Majorana mass term for the right-chiral neutrinos after spontaneous breaking of
lepton number symmetry.  
In keeping with our basic SBCS hypothesis of softly broken
conformal symmetry, that is,  the absence of large intermediate scales between the 
TeV scale and the Planck scale $\MPL$, this mass is here assumed to be 
of $\cO(1\, \TeV)$.  To ensure the stability of the electroweak scale it makes 
use of a novel mechanism to cancel quadratic divergences \cite{CLMN}, relying on the assumed 
existence of a Planck scale finite theory of quantum gravity, as a consequence of which
the cutoff is a physical scale that is not taken to infinity. The phase of the new scalar 
is a Goldstone boson that within the framework of ordinary quantum field theory remains 
massless to all orders due to the vanishing $(B-L)$ anomaly, but will be assumed to
acquire a tiny mass by a quantum gravity induced mechanism, as a result of which
it acquires also small and calculable non-derivative couplings to SM matter. The viability of the 
model  up to the Planck scale will be ensured 
by imposing the consistency requirements listed above.
In particular, the extra degrees of freedom that the CSM contains beyond the
SM are essential for stability: without these extra degrees of freedom the SM does suffer from 
an instability (or rather, meta-stability) because the running scalar self-coupling becomes negative
around $10^{10} \GeV$ \cite{Buttazzo:2013uya}. 

The field content of the model is thus almost the same as for the SM (see {\em e.g.}  
\cite{WeinT2,EPP,Pok} for further details, and \cite{Ramond1} for a more recent update). 
For the fermions we will mostly use SL(2,$\mathbb{C}$) Weyl spinors $\chi_\al$ 
in this paper, together with their complex conjugates $\bar{\chi}_{\dot{\al}}$, see
{\em e.g.} \cite{BaggerWess} for an introduction.  The quark and lepton $SU(2)_L$ 
doublets are thus each composed of two SL(2,$\mathbb{C})$ spinors
\beq\nn 
Q^i \equiv \left(\begin{array}{c} u^i_\al\\[1mm]
                                  d^i_\al
\end{array}\right)\;,\quad
 L^i  \equiv \left(\begin{array}{c} \nu^i_\al\\[1mm]
                                   e^i_\al \end{array}\right)\,,
\eeq
where indices $i,j,...=1,2,3$ label the three families. In addition
we have their $SU(2)_L$-singlet partners $U^{j}_\al$, $D^{j}_\al$ 
and $E^{j}_\al$.  The new fermions in addition to the ones present in the SM
are made up of a  family triplet $N^i_\al$ of gauge singlet neutrinos. 
The scalar sector of the model consists of the usual electroweak scalar doublet 
$H \equiv (H_1, H_2)^\rmt$ and a new gauge-sterile complex singlet scalar $\phi(x)$, 
which carries lepton number. This field couples only to the sterile neutrinos and, via 
the `Higgs portal', to the electroweak doublet $H$.

\subsection{Lagrangian} \label{Sec:Lagrangian}

Apart from the SM-like BRST-exact terms required for gauge fixing \cite{WeinT2}, 
the CSM Lagrangian takes the form
\beq \label{Eq:Lagr-tot}
\cL _{CSM} = \cL_{kin} + \cL_Y - V  \,,
\eeq
with gauge invariant kinetic terms 
\bea\label{Eq:Lagr-kin} 
\cL_{kin} &=& \cL_{kin}^{SM} \,+\,
(\cD_\mu H)^\dagger \cD^\mu\! H   \, +\nn\\[2mm]
&{}& \hspace{-5mm}
+\, \partial_\mu \phi^\star \partial^\mu \phi \,+\,
i \bar{N}^{j}_{\dot{\al}}\overline{\si}^{\mu\,\dot{\al}\be}\partial_\mu{N}^{j}_{\be}\,,
\eea
where we only display the kinetic term of the Higgs doublet and the kinetic terms
of the new fields, while $\cL_{kin}^{SM}$ takes the standard form that can be 
found in any textbook, see \cite{WeinT2,EPP,Pok}. The scalar potential reads
\bea\label{Pot}
V&=&\!\!- \, m_1^2 H^\dagger H-m_2^2 \phi^\star\phi \, +
\phantom{aaaaaaaaaaaaa}
\\[2mm]&{}&\!\!
+\lambda_1(H^\dagger H)^2+2\lambda_3 H^\dagger H\,\phi^\star\phi
+\lambda_2(\phi^\star\phi)^2        \, ,  \nonumber
\eea
with $m_1^2 \, ,\, m_2^2 > 0$.
Exploiting the symmetries of the action we assume that the vacuum expectation 
values take the form\footnote{For the vacuum expectation values we adopt 
  the normalization conventions of \cite{CLMN}.}
\eq{\label{Eq:VEVs}
\sqrt2\langle H_i\rangle=v_H\delta_{i2},
\qquad
\sqrt2\langle\phi\rangle=v_\phi,
} 
with non-negative $v_H$ and $v_\phi$. 
Clearly, we are interested in a situation in which both the electroweak symmetry and lepton
number symmetry are broken, and therefore we assume that $v_H$ and $v_\phi$ 
are non-zero. The values \refer{Eq:VEVs} correspond to the stationary point of \refer{Pot}, 
provided that the mass parameters are chosen as follows
\begin{displaymath} 
m_1^2=\lambda_{3}v_{\phi}^{2}+\lambda_{1}v_{H}^{2}\,,\quad\qquad
m_2^2=\lambda_{3}v_{H}^{2}+\lambda_{2}v_{\phi}^{2}\,.
\end{displaymath}
The tree-level potential \refer{Pot} is bounded from below provided that the quartic couplings obey 
\eq{\label{Eq:PosCond}
\lambda_1>0,\quad \lambda_2>0,\quad 
{\rm and} \quad
\lambda_3 > - \sqrt{\lambda_1 \lambda_2}\,.
}
If, in addition to \refer{Eq:PosCond}, $\lambda_3 < \sqrt{\lambda_1 \lambda_2}$, then \refer{Eq:VEVs} is the global minimum of $V$. The physical spin-zero particles are then 
two CP-even scalars $h$ and $\varphi$, and one CP-odd scalar $a=\sqrt{2}\, \rm{Im}(\phi)$.
 \footnote{Here we employ a linear  parametrization of the scalar fields, {\em i.e.}
$\phi = {\rm Re} \, \phi + i \,{\rm Im} \, \phi$, because this is the most 
convenient one for loop calculations. Later, however, we will switch to an exponential 
parametrization, see Eq. \refer{A} below, which is more convenient to study properties 
of the Goldstone boson, but where the renormalizability of the model is no longer
manifest.} The latter is the Goldstone boson, which -- as we will argue later --
acquires a small mass due to quantum gravity effects (see Sec. \ref{Sec:Break-B-min-L}). 

The two heavy scalar bosons are thus described as mixtures of the 
two real scalar fields with non-vanishing vacuum expectation 
values ($s_\be\equiv \sin\be$, $c_\be\equiv \cos\be$), 
\bea
\left(\begin{array}{c}h\\[1mm] \varphi\end{array}\right)=
\left(\begin{array}{cc}c_\beta&s_\beta\\-s_\beta&c_\beta\end{array}\right)
\left(\begin{array}{c}\sqrt2~\!{\rm Re}(H_2-\langle H_2\rangle)\ \\[1mm]
\sqrt2~\!{\rm Re}(\phi-\langle\phi\rangle)\end{array}\right),\label{eqn:Mixing}
\eea
with masses $M_h$ and $M_\varphi$. The angle $\be$ thus measures the 
mixing between the SM Higgs boson and the new scalar. In order not to be in
conflict with existing data the angle $\be$ must obviously be chosen small,
and furthermore such that $h$ can be identified with the observed SM-like Higgs 
boson with $M_h=(125.6\pm 0.4)\,\GeV$ \cite{PDG2014}. 
Introducing the tree-level SM quartic coupling 
\eq{
\la_0\equiv\frac{1}{2}\frac{M_h^2}{v_H^2}\approx 0.13\,,
} 
one can conveniently parametrize the tree level values of unknown 
parameters $v_\phi$, $M_{\vp}$ and $\be$ in terms of the five parameters
$(v_H, \la_0, \la_1, \la_2, \la_3)$ as follows
\eqs{\label{Eq:vphi-Mphi}
v_\phi
=v_H \sqrt{\frac{\la_0(\la_1 - \la_0)}{\la_2(\la_1 - \la_0)   -  \la_3^2  }   }\,,\;
M_\cphi^2
=2\,\frac{ {\la_1}\la_2-\la_3^2}{\la_0} \,  v_\phi^2
}
and
\eqs{\label{Eq:beta-ang}
\tan\be
=\frac{\la_0-\la_1}{\la_3}\,\frac{v_H}{v_\phi}\,,
\qquad\ \ \ 
s_\be\equiv + \frac{\tan\be}{\sqrt{1+\tan^2\be}}\,.
}
As a consequence, the model predicts the appearance of  a `heavy brother' of 
the usual Higgs boson, which would manifest itself as a {\em narrow} resonance in
or below the TeV region (see Table \ref{Tab:Points} below; the narrowness of the resonance 
is due to the small mixing $\sin^2\beta$ and the relatively large $v_\phi$ scale). The SM-like Higgs 
boson $h$ can, in principle, decay into a pair of pseudo-Goldstone bosons. The corresponding 
branching ratios are, however, very small (for all exemplary points in Table \ref{Tab:Points} 
they do not exceed $0.2\%$). Thus, the decay width of $h$ is decreased with respect to 
the SM value by a factor $\cos^2 \be$. In the numerical analysis we assume that $|\be|\leq 0.3\,$; we note that available LHC data 
 leave enough room for such a modification of SM physics \cite{PDG2016,ATLAS}.

The Yukawa couplings are given by
\bea\label{LY}
\cL_{\rm Y}  &=&
\Big\{
- \, Y_{ji}^D D^{j\al} {H}^{\dagger}\!{Q}^{i}_\al
\,+\,  Y_{ji}^U U^{j\al} {H}^\rmt\!\ep{Q}^{i}_\al \, 
\nn\\[2mm]
&& 
\ \ \, -\, Y_{ji}^E E^{j\al} {H^\dagger}\!{L}^{i}_\al 
 \,  +\,  Y_{ji}^\nu N^{j\al} {H}^\rmt\!\ep{L}^{i}_\al \,  \nnn[2mm]
&& \ \ \,  - \, \frac{1}{2}Y^M_{ji} \phi N^{j\al} N^i_\al\Big\} \, +\, {\rm h.c.}
\,,
\qquad
\eea
with the usual Yukawa matrices $Y^E, Y^D$ and $Y^U$ of the SM,  where $\ep$ is the 
antisymmetric  $SU(2)_L$ metric. The matrix $Y^\nu$ mediates the coupling of the SM
fields to the sterile neutrino components, while the complex symmetric matrix $Y^M$ 
describes the interactions of the latter with the new scalar $\phi$. 
By fermionic field redefinitions that preserve Eq. \refer{Eq:Lagr-kin},  one can assume that 
$Y^M$, $Y^E$ and $Y^U$ are diagonal and non-negative, and that $Y^D$ differs from a 
positive and diagonal $\tilde{Y}^D$ by the (inverse of) unitary CKM matrix $Y^D=\tilde{Y}^D V_{CKM}^{\ \dagger}$. 
In more customary notation the fermionic fields are described by 4-component Dirac 
spinors of charged leptons and up-type quarks
\eq{\label{Eq:Dirac-Field}
\Psi^{i}_{E}
=
\left[
\begin{array}{l}
e^{i}_\al\\[2mm]
\bar{E}^{i\dot{\al}}\\
\end{array}
\right]
,
\qquad\ 
\Psi^{i}_{U}
=
\left[
\begin{array}{l}
u^{i}_\al\\[2mm]
\bar{U}^{i\dot{\al}}\\
\end{array}
\right]
,
}
together with the analogous 4-spinor field $\Psi^{i}_{D}$ 
\beq\label{Dirac2}
\Psi^{i}_{D}
=
\left[
\begin{array}{l}
d'^{\, i}_\al\\[2mm]
\bar{D}^{i\dot{\al}}\\
\end{array}
\right]
\eeq
for the down quarks, with a $V_{CKM}$-induced rotation $d^i_\al \rightarrow d'^{\,i}_\al$
of the upper components \cite{WeinT2}. 
\footnote{The {\em chiral} Dirac fields usually employed are thus 
$e_L\equiv P_L\Psi_E = (e,0)$ and $e_R\equiv P_R\Psi_E = (0,\bar{E})$, {\em etc.}.} 

After spontaneous symmetry breaking the neutrino mass terms are
\bea
-\cL &\supset &  
\, m_{D{ij}}\, N^{\al i} \nu^j_\al \,+\, {m}_{D{ij}}^\star\, \bar{N}^i_{\dot\al} \bar{\nu}^{\dot\al j} \nnn[2mm]
&&\,+\,
\frac{1}{2}M_{N{ij}}\, N^{i\al} N^j_\al \,+\, \frac{1}{2} M_{N{ij}}^{\star}\, \bar{N}^i_{\dot\al} \bar{N}^{\dot\al j}   
\eea
with 
\eq{
M_N\equiv Y^M\, v_\phi/\sqrt{2}\, ,
\qquad\quad 
m_D \equiv Y^\nu\, v_H/\sqrt{2}\,.
}
The masses for light neutrinos are thus obtained via the  seesaw mechanism \cite{Min,seesaw,Yan} and follow easily by diagonalizing the symmetric tree-level
mass matrix (see also \cite{Latosinski})
\kM\beq   
\cM \,=\,
\left[
\begin{array}{cc}
 0 & {m_D}^\rmt \\[2mm]
{m_D} & M_N \\
\end{array}
\right]\,,
\label{combination}
\eeq

Introducing unitary $3\times3$ matrices $U_0$ and $V_0$ and the $6\times 6$ block matrix 
\eq{\label{Eq:bV}
\bV=\left[\begin{array}{cc}
\bX_1 & \bX_2\\
\bX_3 & \bX_4
\end{array}\right]\,,
}
with the $3\times 3$ submatrices
\eqs{
\bX_1&=&i\Big\{\id -\frac{1}{2}\,m_D^\dagger M_N^{-1\dagger}\, M_N^{-1}\,m_D\Big\} U_0\,,\nn\\
\bX_2&=&m_D^\dagger\,M_N^{-1\dagger}\, V_0\,,\nn\\
\bX_3&=&-i\,M_N^{-1}\,m_D\, U_0\,,\nn\\
\bX_4&=&\Big\{\id -\frac{1}{2}M_N^{-1}\,m_D\,m_D^\dagger\, M_N^{-1\dagger}\Big\} V_0\,,\nn
}
one has $\bV^\dagger \bV=\id + \cO(|\!|m_D|\!|^3)$, and
\eq{\label{Eq:DiagMas}
\cM_{\rm ph}\equiv \bV^\rmt\!\cM\,\bV=
\left[\!
\begin{array}{cc}
U_0^\rmt\, \cM_\nu\, U_0  & 0\\
0 & \!\!
V_0^\rmt\, \cM_N\, V_0 
\end{array}
\!\right]
\!+\!\cO(|\!|m_D|\!|^3)
\,,
}
with complex symmetric matrices
\eqs{
\cM_\nu
&=&m_D^\rmt\,M_N^{-1}\,m_D\,, \\
\cM_N
&=&
M_N
+\frac{1}{2}M_N^{-1\star}\, m_D^\star\,m_D^\rmt 
+\frac{1}{2}m_D\,m_D^\dagger\, M_N^{-1\dagger}
 \,.\quad\label{Eq:cM_N} 
}
Observe that up to $\cO(m_D^3)$ the matrix $\bV$ achieves the diagonalization of the 
$6 \times 6$ matrix
$\cM$ in (\ref{combination}) into the two blocks of $3\times 3$ matrices exhibited above, 
but that the latter are not necessarily in diagonal form yet.
Employing the Casas-Ibarra parametrization \cite{CA-IB} of the Dirac mass matrix $m_D$ 
(or equivalently the Yukawa matrix $Y^\nu$; as explained above, $M_N$ can be assumed
positive diagonal)
\kN\eq{\label{Eq:CaIbPar}
m_D =M_N^{1/2}\,\RCI^\rmt \,
\Big[\diag(m_{\nu 1},\,m_{\nu2},\, m_{\nu3})^{1/2}\Big]
\,U_{MNS}^{\dagger}\,,  
}
with the unitary Maki-Nakagawa-Sakata $U_{MNS}$ matrix (see \cite{CA-IB} and references therein) and a complex orthogonal Casas-Ibarra matrix $\RCI$ ($\RCI^{\ \rmt}\,\RCI^{\phantom{\rmt}}=\id$)
one has 
\eq{
\cM_\nu=U_{MNS}^{\star}\,\Big[\diag(m_{\nu 1},\,m_{\nu2},\, m_{\nu3})\Big]
\,U_{MNS}^{\dagger}\,,  
}
which shows that $m_{\nu\,i}$ are light neutrino masses at the tree level. The main advantage 
of the Casas-Ibarra parametrization, and the reason we use it here, is that it provides a clear 
separation of the parameters of $Y^\nu$ into the ones that are relevant for neutrino oscillations,
namely $m_{\nu\,i}$ and the CKM-like unitary matrix $U_{MNS}$, and the ones describing 
heavy neutrinos and their properties ($M_N,\  \RCI$).

The matrix $V_0$ is now chosen such as to make $V_0^\rmt\, \cM_N\, V_0$ a positive diagonal 
matrix (note that $\cM_N$ differs from $M_N$, cf. (\ref{Eq:cM_N})!). The matrix
$\cM_{\rm ph}$ in Eq. \refer{Eq:DiagMas} is then diagonal provided that 
$U_0=U_{MNS}$; however, we will be mainly interested in light neutrino states that participate in specific fast interactions during the leptogenesis, {\em i.e.} that are approximate eigenstates 
of weak interactions. Therefore, we take $U_0=\id$,  and change the basis in the field space 
so that $\cM_{\rm ph}$ in Eq. \refer{Eq:DiagMas} is a new mass matrix (in other words,
we are using interaction eigenstates rather than mass eigenstates for the light
neutrinos). Henceforth, $\nu^i_\al$ and $N^j_\al$ denote neutrino fields in this new 
basis, unless stated otherwise, and are referred to as light and heavy neutrinos. 
It is sometimes convenient to assemble these 2-component Weyl spinors into  
Majorana 4-spinors 
\eq{\label{Eq:Majorana}
\psi^{i}_{N}
=
\left[
\begin{array}{l}
N^{i}_\al\\[2mm]
\bar{N}^{i\dot{\al}}\\
\end{array}
\right]
,
\qquad\ 
\psi^{i}_{\nu}
=
\left[
\begin{array}{l}
\nu^{i}_\al\\[2mm]
\bar{\nu}^{i\dot{\al}}\\
\end{array}
\right]
\,.
}
Note that, as a result of the rotation with $\bV$, the (new) $N$'s do couple to the 
massive gauge bosons already at the tree-level 
\eqs{\label{Eq:N-to-Z}
\cL_{ ZN\nu} &=& 
i\,Z_\mu\,\Big(
\cF^{(Z)}_{j i} 
\bar{N}^{j}_\da\,\bar{\si}^{\mu\,\da \al}\,{\nu}^{i}_\al
+
\cF^{(Z)\star}_{j i}
{N}^{j\al}{\si}_{\al\da}^\mu\,\bar{\nu}^{i\da}  
\Big)
\nn\\[7 pt]
&\equiv& 
i\,Z_\mu\,\,
\bar\psi_{N}^{j}\,\gamma^\mu 
\Big(  \cF^{(Z)}_{j i}\, P_L+\cF^{(Z)\star}_{j i}\, P_R  \Big)
\psi_{\nu}^{i}\,,
}
\eqs{\label{Eq:N-to-W}
\cL_{ WNe} 
&=&
i\,(W^1_\mu - i W^2_\mu) \,\cF^{(W)}_{j i}\,
\bar{N}^{j}_\da\,\bar{\si}^{\mu\,\da \al}\,{e}^{i}_\al
\,+ \, {\rm h.c.}
\nn\\[7pt] 
&\equiv&
i\,(W^1_\mu - i W^2_\mu) \,\cF^{(W)}_{j i}\,
\bar\psi_{N}^{j}\,\gamma^\mu 
\, P_L\, \Psi_{E}^{i}
\,+ \, {\rm h.c.}
\,,\ \ \ \ 
}
where for clarity we also give the result in standard 4-spinor notation, and 
where $P_{L/R} \equiv \frac12 (1 \mp \gamma^5)$ are the usual chiral projectors.
The matrices $\cF^{(Z,W)}$ follow immediately from Eq. \refer{Eq:bV}
\eqs{
\cF^{(Z)}&=&-\frac{i}{2}(g_w^2+g_y^2)^{1/2}\, \bX_2^\dagger \bX_1\,,
\\[4 pt]
\cF^{(W)}&=&-\frac{i}{2}g_w\, \bX_2^\dagger \,.
}
To avoid confusion we also use calligraphic letters to denote the couplings between 
the new fields $N/\nu$ and the scalars $S=h,\cphi,a$, to wit,
\koment{str.PDCN24}
\eqs{\label{Eq:N-to-h}
\cL_{SN\nu} &=& 
-
S\, 
\left(
\cY^{(S)}_{j i}  N^{j\al}\nu^{i}_{\al}
+
\cY^{(S)\,\star}_{j i} \bar{N}^{j}_{\da}\bar{\nu}^{i\da}
\right) 
\,\nn\\[7 pt]
&\equiv&-
S\, \bar\psi_{N}^{j}
\left(
\cY^{(S)}_{j i}P_L+
\cY^{(S)\,\star}_{j i} P_R\right)
\psi_{\nu}^{i}\,,
}
where the leading terms in $m_D$ read
\kO\eqs{
\cY^{(h)}&=&+i\left\{\frac{c_\be}{v_H}-\frac{s_\be}{v_\phi}\right\}V_0^{\rmt}\,m_D\,U_0\,,
\nn\\[7pt]
\cY^{(\cphi)}&=&-i\left\{\frac{s_\be}{v_H}+\frac{c_\be}{v_\phi}\right\}V_0^{\rmt}\,m_D\,U_0\,,
\nn\\[7pt]
\cY^{(a)}&=&\frac{1}{v_\phi}\,V_0^{\rmt}\,m_D\,U_0\,,\nn
}
(as said, $U_0=\id$).
Because a main postulate behind the CSM is the presumed absence of any intermediate scales 
between the electroweak scale and the Planck scale $\MPL$, the scale 
of lepton number symmetry breaking $v_\phi$ is assumed to lie in the $\TeV$ range.    
With  $Y_M \sim 1$, the masses of heavy neutrinos are relatively small, and the light neutrino data \cite{PDG2014} indicate that $Y^\nu$ is of order $Y^\nu \sim 10^{-6}$. 
To allow for baryon number generation despite the low masses of heavy neutrinos, 
the mechanism of `resonant leptogenesis' was proposed and explored in
 \cite{PilafRL1,RL2,RL3,RL4}. This mechanism is based on the observation that CP-violation 
 (a crucial ingredient in dynamically generated baryon asymmetry \cite{Sakh}) is enhanced 
 whenever the masses of heavy neutrinos are approximately degenerate. 
Accordingly, we assume that the Yukawa Majorana matrix is in fact {\em proportional
to the unit matrix}, that is, 
\beq\label{Eq:YM-deg}
Y^M_{ij} = y_M \delta_{ij}
\eeq
with $y_M \sim \cO(1)$. Consequently there is an approximate SO(3) symmetry in the heavy  
neutrino sector, which is only very weakly broken by the Yukawa couplings $Y^\nu$. For definiteness, we assume Eq. \refer{Eq:YM-deg} to hold at the electroweak scale, for the $\overline{\rm MS}$ renormalization scale $\mu=M_{top}$. In turn, the mass splitting of heavy neutrinos is entirely due to the seesaw mechanism, Eq. \refer{Eq:cM_N}. (As emphasized in \cite{PilafSMM} the SO(3) symmetry ensures that \refer{Eq:YM-deg}  is stable against quantum corrections in a good approximation; nonetheless, when \refer{Eq:YM-deg}
holds instead at high RG scale $\mu_{*}\sim \MPL$, then $Y^{\nu}$-induced RG-splitting of $y_M$'s yields splitting of heavy neutrino masses that is of similar order as the seesaw one, see e.g. \cite{BLJ}). It should be stressed here that, due to the degeneracy \refer{Eq:YM-deg}, the $V_0$ matrix in Eq. \refer{Eq:DiagMas} is clearly \emph{not} an $\cO(Y^\nu)$ perturbation of the identity matrix; this is technically similar to (though physically different from) the Dashen's vacuum realignment condition \cite{DASHEN} (see also \cite{WeinT2}).

\subsection{Cancelling Quadratic Divergences} \label{Sec:QuadDiv}

We stress again the presence of explicit scalar mass terms in (\ref{Pot}), in contrast to the original 
model of \cite{KM} which relied on the CW mechanism \cite{CW} to break electroweak symmetry. 
Our main reason for this is that the CW mechanism does not eliminate quadratic divergences, 
and thus the low energy theory would remain sensitive to Planck scale corrections. 

At one loop the coefficients of the quadratic divergences $\La^2$ for the two scalar fields are \cite{CLMN}
\bea\label{quadDiv}
16\pi^2f^{\rm quad}_1(\la,g,y)&=&6\la_1+2\la_3+\frac{9}{4}g_w^2
+\frac{3}{4}g_y^2-6y_t^2\nn\\[2mm]
16\pi^2f^{\rm quad}_2(\la,g,y)&=&4\la_2+4\la_3 -  3 y^2_{{M}}.
\label{eqn:fconds}\phantom{aaaaaaaa}
\eea
Here $g_w$ and $g_y$ are the $SU(2)_{L}\times U(1)_Y$ gauge couplings, while
$y_t$ is the top quark Yukawa coupling. 
For simplicity (and without much loss in precision) we neglect all other
Yukawa couplings. Note that Eqs. \refer{quadDiv} are independent of the details of the cutoff regularization, as long as the regulator (here assumed to be provided by the quantum theory of gravity) acts in the same way on all fields. Of course, another crucial assumption here is that 
we can neglect contributions of graviton loops to \refer{quadDiv}; this assumption is based on the hypothesis that the UV finite theory of quantum gravity effectively screens these contributions from low energy physics. 

An obvious question at this point is the following. One would at first think that 
Eqs. \refer{quadDiv} depend on the renormalization scale $\mu$ via the RG running of 
the couplings, a well-known issue in the context of Veltman's conditions \cite{Veltman}.
This is, however, only apparent, since when all higher corrections are included, 
the functions $f_1,f_2$ obey appropriate renormalization group equations, in such a way
that the implicit scale dependence is exactly canceled by the explicit presence of $\log(\mu)$ introduced by higher loop corrections. Therefore, the all-order coefficients  $f_1,f_2$ are in fact
$\mu$-independent (and $\La$-independent) functions of the bare couplings $\la_B$ (which 
themselves  depend on the cutoff $\La$, as the latter is varied). Thus, the couplings appearing on the 
right-hand-side of \refer{quadDiv} are $\la_B(\La)$ {\em etc.}, rather than 
$\la(\mu)|_{\mu=M_{top}}$.\footnote{In writing these equations
we suppress a reference scale $\mu_0$ needed to render the arguments dimensionless.
The latter can be chosen as $\mu_0 = M_{top}$, or alternatively as $\mu_0 \equiv \mu$, in
which case all couplings would depend only on the ratio $\mu/\Lambda$ where 
the cutoff $\Lambda$ is kept fixed (which is the case we consider below).}
Nonetheless, employing running couplings $\la(\mu)$ is convenient also in the present 
context, as these allow for a resummation of leading logarithms in the  relation between the bare 
couplings $\la_B(\La)$ and the renormalized ones $\la_R=\la(\mu)|_{\mu=M_{top}}$, via the 
usual renormalization group improvement (see e.g. \cite{Kastening:1991gv,Ford:1992mv,MN}). 
In fact, in a minimal-subtraction-type scheme based on cutoff regularization \cite{CLMN} 
(below called $\La$-MS), the bare couplings $\la_B(\La)$ coincide with the running 
couplings $\la(\mu)$ corresponding to $\mu=\La$,
\eq{
\la_B(\La) \equiv  \la(\mu)\big|_{\mu=\La}\,,
}
see also \cite{ChLM} for a discussion of the issues appearing in cutoff regularized gauge theories. 

The appearance of bare couplings in \refer{quadDiv} can also be motivated and understood
from the point of view of constructive QFT (see e.g. \cite{Glimm}), although we are,
of course, aware that there is no rigorous construction of the SM.
There one attempts to rigorously construct a functional measure
for interacting QFTs. This requires the introduction of both UV and IR ({\it i.e.} finite
volume) regulators. For the regularized theory one then introduces counterterms 
as functions of the {\em bare} parameters $\la_B(\La)$ and tries to adjust the latter
as functions of the  UV cutoff $\Lambda$ in such a way that the theory gives well defined
physical answers in the limit $\Lambda\rightarrow\infty$ (in which the bare couplings
usually assume singular values).
In particular, for a given value of the cutoff one can thus impose the vanishing of
the coefficient of the quadratic divergence as a single condition on the bare parameters. 
In that framework running couplings $\la(\mu)$ play no role; they are merely an auxiliary
device to conveniently parametrize the scale dependence of correlation functions. 

In summary, the coefficients of quadratic divergences \refer{quadDiv} are \emph{calculable} 
functions of the cutoff scale $\La$, provided that all low energy parameters
$\la_1(\mu)|_{\mu=M_{top}}$ {\em etc.} are fixed by experiment. To determine the 
evolution of the couplings from $\mu=M_{top}$ up to $\Lambda$ (where they are identified
with the bare couplings) in the leading logarithmic (LL) 
approximation we need only the one-loop beta functions \cite{CLMN} 
(we use the notation $\tilde\beta\equiv 16\pi^2\beta$; furthermore we make use
of \refer{Eq:YM-deg})

\bea
\tilde\beta_{\la_1}&=&24\la_1^2+4\la_3^2
-3\la_1\left(3g_w^2+g_y^2-4y_t^2\right)
\nn\\
&{}&
+{9\over8}g_w^4+{3\over4}g_w^2 g_y^2+{3\over8}g_y^4- 6 y_t^4  \nn\\[7 pt]
\tilde\beta_{\la_2}&=& 20\la_2^2+8\la_3^2
+6\la_2 y_{M}^2-3y_{M}^4  \nn\\[7 pt]
\tilde\beta_{\la_3} &=&
\frac{1}{2}\la_3\Big\{24\la_1+16\la_2+16\la_3
\\
&& \ \ \ \ \  \ \
- \left(9g_w^2+3g_y^2\right)
+6y_{M}^2+12y_t^2 \Big\} \, \nn
\eea
\bea
\tilde\beta_{g_w} &=& -\frac{19}{6}g_w^3\;, \ \
\tilde\beta_{g_y}=\frac{41}{6}g_y^3, \ \
\tilde\beta_{g_s} =-7 g_s^3,\nn\\[7 pt]
\tilde\beta_{y_t} &=& y_t\left\{ \frac{9}{2}y_t^2
-8g_s^2  -\frac{9}{4}g_w^2
-\frac{17}{12}g_y^2 \right\}, \nn\\[7 pt]
\tilde\beta_{y_{{M}}} &=& \frac{5}{2}{y_{{M}}}^3
\,,
\eea
which show in particular how the $SU(3)_c$ gauge coupling $g_s$ affects the evolution of $y_t$
so no Landau pole develops for $y_t$. This effect is also seen in the other expressions
where bosonic and fermionic contributions balance each other in such a
way that the theory remains perturbatively under control up to $\MPL$ 
(with appropriate initial values).   

At this point it should be stressed that all the ingredients necessary to find the coefficients 
$f_1,f_2$ with resummed next-to-leading logarithms are at our disposal. In particular, 
the two-loop beta functions in $\La$-MS together with the two-loop 
coefficients in a generic renormalizable model are given in \cite{ChLM} (one can also find there the generic one-loop relation between renormalized parameters in $\La$-MS and their counterparts in the conventional $\overline{\rm MS}$ scheme of dimensional regularization). However, as most of the parameters of CSM are still unknown, we are content here with resummation of the leading logarithms only. The rationale behind this restriction, is that the \emph{one-loop} RG evolution in gauge-Yukawa sector is independent of quartic scalar couplings, which significantly simplifies the scan over the parameter space; in particular $y_t$ and gauge couplings at the Planck scale are known. Recall that 
the one-loop beta functions reflect the structure of non-local terms in one-particle-irreducible effective action $\Ga[\cdot]$, and thus are universal across different regularizations (at least in the class of mass independent renormalization schemes \cite{Weinberg:MassIndepScheme}, 
to which $\La$-MS belongs).  Therefore the RG-improved coefficients \refer{quadDiv} at the LL order are independent of the details of cutoff regularization as well.

We note that the cutoff dependence of coefficients of quadratic divergences in 
the pure SM was already analyzed in \cite{HKO} where it was found that they cancel for
$\La\approx 10^{24}\,\GeV)$, and thus (logarithmically speaking) not so far from the Planck scale. 
This observation  motivated our proposal that the vanishing of quadratic divergences
at the Planck scale, and thus stabilization of the electroweak scale, may be achieved
by means of a `small' modification of the SM like the one proposed here. Ultimately,
the cancellation of quadratic divergences would be  due to a still unknown quantum
gravity induced mechanism which is different from low energy supersymmetry (but which could
still involve Planck scale supersymmetry in an essential way). 

From the perspective of effective field theory (EFT), valid for energies $E\lesssim\MPL$, we have a clear distinction between low energy ($N=1$) supersymmetry and the present proposal. 
In supersymmetric models, the underlying mechanism of quantum gravity appears via
the  (super)symmetry of the EFT itself, and thus the cancellation holds independently 
of the value of the cutoff
\eq{\nn
f^{\rm quad}_{\rm SUSY}(\La)= 0\,, \quad\forall\La\,.
} 
By contrast, in the present context the absence of quadratic divergences (and thus
the stabilization of the electroweak scale) manifests 
itself via the existence of a distinguished value $\La_*$ of the cutoff (close 
to the Planck scale) such that $f(\La_*)=0$. Importantly, the question whether 
or not such a scale exists for which both coefficients \refer{quadDiv} vanish, can 
in principle be answered provided that all CSM parameters can be 
measured with sufficient accuracy.

%

We therefore assume that such a distinguished value close to $\MPL$ exists, 
so we can impose the conditions
\beq\label{quadDiv1}
f_1^{\rm quad} (\la,g,y) \,=\,f_2^{\rm quad} (\la,g,y) = 0
\eeq
on the running couplings with $\mu$ equal to the (reduced) Planck scale; from a 
low-energy perspective these can be considered as an RG-improved version
of Veltman's conditions \cite{Veltman}.  Disregarding
the other SM couplings this condition restricts the four-dimensional space of parameters 
$(\la_1 , \la_2 , \la_3 ,y_M)$, cf. Eqs. \refer{Eq:vphi-Mphi}, to a two-dimensional 
submanifold.\footnote{Because the conditions \refer{quadDiv1} are RG-invariant, our 
 approach bears also some resemblance to Zimmerman's reduction of couplings \cite{Sibold1,Sibold2}.}
To implement our conditions in practice we then evolve the couplings 
along this submanifold
from $\MPL$ back down to the electroweak scale $\mu=M_{top}$ and calculate the 
masses and mixing angle using Eqs. \refer{Eq:vphi-Mphi}--\refer{Eq:beta-ang}. 
Moreover, to ensure  perturbativity we demand that all running couplings (including $y_M$) 
remain small over the whole range of energies between $M_{top}$ and $\MPL$ 
(more concretely, for our numerical checks we demand $0<\la_1,\la_2,y_M < 2$, and $-2<\la_3 < 2$, see also the next subsection; in practice for all points in Table \ref{Tab:Points} scalar self-couplings at the electroweak scale are smaller than 0.25). It should be 
stressed that this approach is consistent because the values of the gauge and Yukawa 
couplings at the Planck scale are independent of the values of quartic couplings, 
as far as leading logarithms are concerned.

\subsection{Stability of electroweak vacuum}

One of the very few `weak spots' of the pure SM is the meta-stability  of the electroweak 
vacuum \cite{Buttazzo:2013uya}.  Namely, the effective potential of the SM (with appropriately resummed large logarithms) develops a new deeper minimum for 
$H\gtrsim 10^{10}\,\GeV$, thus implying an instability of the  electroweak vacuum 
via quantum mechanical tunneling. This can be seen also more heuristically, by following 
the RG evolution $\la = \la(\mu)$ of the scalar self-coupling and noticing that for 
$\mu \sim 10^{10}\,\GeV$ the function $\la(\mu)$ dips below zero due to 
the large negative contribution from the top quark \cite{Buttazzo:2013uya} 
(but becomes positive again for yet larger values of $\mu$). For values of the fields that 
are much larger than the electroweak scale, the full effective potential of the SM is well 
approximated by the quartic term
\eq{
V_{\rm eff}(H)\, \approx  \tilde{\la}\,  (H^\dagger H)^2
\,, 
}
However, here one cannot simply substitute the self-coupling at the electroweak scale;
rather, in order to avoid huge logarithmic corrections on the right-hand-side, the 
correct value of the quartic coupling $\tilde{\la}$ in the above formula is obtained by
substituting the running coupling
evaluated at the appropriate energy scale of the order of $\sqrt{H^\dagger H}$, {\em i.e.} 
\eq{
\tilde\la=\la(\mu)|_{\mu\equiv\sqrt{H^\dagger H}}\,,
}
rather than $\tilde\la=\la(\mu)|_{\mu=M_{\rm top}}$. For theories like the SM, in which the effective potential depends only on a single field (up to the orbits of symmetry group), this somewhat heuristic reasoning can be put on firmer grounds, by resumming large logarithms via the 
renormalization group improvement \cite{Kastening:1991gv,Ford:1992mv,MN}.

For the CSM there are now {\em two} scalar fields (up to symmetries of $V_{\rm eff}$) and the situation is more complicated, basically because with more than
one scalar field, the RG-improvement cannot simultaneously determine
the resummation of logarithms in all directions in field space.
For this reason we have to rely on the more heuristic argument, by demanding
that the positivity conditions \refer{Eq:PosCond} be satisfied not only at the electroweak 
scale $\mu=M_{\rm top}$, but also for the running couplings at all intermediate scales 
$M_{top} < \mu < \MPL$. This provides a strong indication that the electroweak vacuum \refer{Eq:VEVs} in the CSM remains the global minimum of the full effective potential, at 
least in the region $|\phi|^2<\MPL^2$, $H^\dagger H<\MPL^2$, in which EFT is valid. 
Thus, following the RG evolution from the Planck scale, where the conditions 
\refer{quadDiv1} are imposed, down to the electroweak  scale we impose the inequalities 
\beq
\la_1(\mu) > 0\;, \quad\la_2(\mu) > 0\;, \quad
\la_3(\mu) >- \sqrt{ \la_1(\mu) \la_2(\mu) }\,.
\eeq
{\em in addition} to the conditions enunciated at the end of the foregoing subsection.
These extra stability conditions lead to further restrictions on the parameters.
It is therefore  a non-trivial fact that parameter ranges exist which satisfy all
these conditions and restrictions.

A set of exemplary points consistent with all our restrictions is given in Table \ref{Tab:Points}. $\Ga_{h,\cphi}$ denote decay width of the Higgs particle $h$ and its `heavy brother' $\cphi$.  
Br($\cphi\to$[SM]) is the branching ratio for SM-like decay channels of $\cphi$, while non-SM-like decay channels of $h$ are negligible for all points in the Table. $\sY_{B0}$ denotes the current 
baryon number density to entropy density ratio calculated on assumptions 
specified in Sec. \ref{Sec:Lepto}. In particular, the Table displays a viable range of
mass values for both the new scalar and the heavy neutrinos. For all points the heavy neutrinos are heavier than the new scalar field $\cphi$, and thus their decays are the main source of lepton asymmetry.  Note also the relatively large 
values of $v_\phi$ which are necessary for successful leptogenesis. This comes about
because $y_M$ must remain sufficiently small
so as to allow for the departure of heavy neutrinos from thermal equilibrium, while their masses $\sim y_M v_\phi$ should be large enough so that the departure takes place when baryon-number violating processes are still fast. 
Importantly, the values of dimensionless couplings corresponding to all points in the Table are small while masses of new states are comparable to the electroweak scale; thus one can trust that radiative corrections to the tree-level masses {\em etc.} are small.

\onecolumngrid
\begin{center}
\begin{table}[t]
\caption{Exemplary values} \label{Tab:Points} 
\vspace{3mm}
\begin{tabular}{|c|c|c|c|c|c|c|c|c|}
\hline
$M_\cphi[\GeV]$  & $\ \ \ s_\be\ \ \ $ & $M_N[\GeV]$ & $v_\phi[\GeV]$ & 
$\ \ \ \ \ \ \  \sY_{B0}\ \ \ \ \ \ \ $   & $\Gamma_h[\MeV]$  & $\Gamma_\varphi[\GeV]$ &
   Br($\varphi \to$ [SM])  & 
   Br($\varphi \to hh $) \\\hline\hline
 1030 & -0.067 & 1604 & 17090 &
  7.9 $\times 10^{-11}$ & 4.19 &
   4.02 & 0.78 & 0.2 \\\hline
 893 & -0.076 & 1238 & 11331 &
   1.2$\times10^{-10}$ & 4.186 &
   3.3 & 0.76 & 0.2 \\ \hline
 839 & -0.082 & 1181 & 11056 &
   1.2$\times10^{-10}$ & 4.182 &
   3.08 & 0.76 & 0.2 \\\hline
 738 & -0.093 & 1052 & 10082 &
   1.1$\times10^{-10}$ & 4.174 &
   2.66 & 0.76 & 0.22 \\\hline
 642 & -0.11 & 1303 & 19467 &
   1.7$\times10^{-10}$ & 4.16 &
   2.34 & 0.76 & 0.22 \\\hline
 531 & -0.13 & 949 & 12358 &
   1.5$\times10^{-10}$ & 4.138 &
   1.92 & 0.74 & 0.22 \\\hline
 393 & -0.18 & 801 & 12591 &
   1.0$\times10^{-10}$ & 4.07 &
   1.28 & 0.72 & 0.26 \\\hline
 362 & -0.20 & 815 & 14534 &
   1.6$\times10^{-10}$ & 4.04 &
   1.06 & 0.68 & 0.3 \\\hline
 350 & -0.21 & 738 & 12302 &
   7.4$\times10^{-11}$ & 4.028 &
   0.96 & 0.66 & 0.32 \\\hline
 320 & -0.23 & 751 & 14437 &
   1.3$\times10^{-10}$ & 3.984 &
   0.86 & 0.66 & 0.32 \\\hline
 279 & -0.28 & 683 & 14334 &
   9.6$\times10^{-11}$ & 3.896 &
   0.68 & 0.7 & 0.28 \\\hline
 258 & -0.31 & 675 & 15752 &
   1.3$\times10^{-10}$ & 3.824 &
   0.54 & 0.78 & 0.20 \\\hline
\end{tabular}
\end{table}

\end{center}
\twocolumngrid

\section{Resonant Leptogenesis}\label{Sec:Lepto}

By assumption the lepton number symmetry $L$ of the CSM is spontaneously 
broken by the non-vanishing vacuum expectation value  $\langle\phi\rangle\sim \cO(1\,\TeV)$. 
The proper quantity to study is therefore the  lepton number density $L'$ of the SM
under which heavy neutrinos have vanishing charges.\footnote{In
  the following we adopt the convention that primed quantities refer to the pure SM, while unprimed
  letters refer to the CSM with its enlarged set of fields. For instance, $L'_i$ is the lepton
  number of species $i$ in the pure SM, which thus excludes the contributions of $N_i$
  and the new scalar $\phi$.} The individual lepton number 
symmetries $L'_i$,  $i=e,\mu,\tau\,,$ of the SM (with $L'=\sum_iL'_i$) are only weakly 
broken by $Y^\nu$-effects as well as by gauge anomalies. 

In the framework of leptogenesis \cite{YanaFuku} the baryon number density $n_B$ in 
the universe \cite{Planck,PDG2014}\footnote{This number is often given by normalizing
 with respect to the entropy density, see (\ref{Yscr}) and (\ref{YB0}) below.}
\eq{\label{Eq:nB}
{n_B} = (6.05 \pm 0.07) \times 10^{-10}\, n_\gamma \,,
}
(where $n_\gamma$ denotes the number density of photons)
is produced by non-perturbative SM interactions that break baryon and lepton number 
symmetries down to the non-anomalous combination $(B-L')$, and generate baryons 
from non-vanishing lepton number density $n_{L'}$ via the usual sphaleron mechanism \cite{SFAL}. 
Thus the problem can be reduced to that of explaining the lepton asymmetry $n_{L'}$,
which itself is produced in lepton number and CP violating out-of-equilibrium 
decays of heavy neutrinos, as they occur in the CSM. In this way
all the Sakharov conditions \cite{Sakh} can be satisfied. \footnote{See also
 \cite{BLJ} for a discussion of resonant leptogenesis for a CSM-like model with
 {\em gauged} $(B-L)$ symmetry.}

To achieve the correct order of CP-violation despite small $Y^\nu$ values, we rely on 
the mechanism of ``resonant leptogenesis" \cite{PilafRL1,RL2,RL3,RL4}, which 
can be naturally realized within the present scheme as a consequence of the 
assumed degeneracy of the Yukawa matrix $Y^M$, cf. Eq. \refer{Eq:YM-deg}. 
The baryon number density ${n_B}$ can then be calculated by solving the relevant 
Boltzmann equations (see {\em e.g.} \cite{Kolb-Turner-Book}).

\subsection{CP-violation}

The CP-asymmetries relevant for calculation of $n_B$ are 
\eqs{\label{Eq:ep-def-nowa}
\vep_{j i}^{(h\nu)}
&\equiv&
\frac{\Gamma(N_{j}\to h\nu_{i})-\Gamma(N_{j}\to h\bar\nu_{i})}
     {\Gamma(N_{j}\to h\nu_{i})+\Gamma(N_{j}\to h\bar\nu_{i})}\,,
\\[10 pt]
\label{Eq:ep-def-nowa-Z}
\vep_{j i}^{(Z\nu)}
&\equiv&
\frac{\Gamma(N_{j}\to Z\nu_{i})-\Gamma(N_{j}\to Z\bar\nu_{i})}
     {\Gamma(N_{j}\to Z\nu_{i})+\Gamma(N_{j}\to Z\bar\nu_{i})}\,,
}
together with their counterparts with additional scalars (or $W$-bosons and charged leptons) 
in the final states.  The tree-level contributions to the decay widths in the formulae above 
follow immediately from the vertices in Eqs. \refer{Eq:N-to-h} 
and \refer{Eq:N-to-Z}-\refer{Eq:N-to-W}, 
\footnote{Here we can neglect the light neutrino masses in very good approximation. Thus 
 the field $P_L\psi_\nu\simeq \nu_\al$ in \refer{Eq:N-to-h} annihilates neutrinos $\nu$ 
 and creates antineutrinos $\bar\nu$, while $P_R\psi_\nu\simeq \bar\nu^\da$ does the 
 opposite, see also the appendix for an explicit description of the neutrino operators
 in the SL(2,$\mathbb{C}$) basis. }
while non-zero contributions to $\vep_{ji}$ originate from the interference between these tree-level vertices and loop diagrams describing the correction to proper vertices and external lines \cite{LS}. Generically,  both kinds of corrections are of the same order \cite{LS}, and are way too small to ensure a successful leptogenesis for $Y^\nu$ having the matrix elements of the order of 1$0^{-6}$. However, the external line corrections are resonantly enhanced for (approximately) degenerate masses of heavy neutrinos \cite{PilafRL1,RL2,RL3,RL4}. 

In fact, calculation of `external' line corrections (especially in the resonant regime) requires some care, since the incoming states correspond to unstable particles. In \cite{PilafUnd}, see also \cite{PilafRL1,PlumOld,PlumNew}, the CP-asymmetry 
$\vep_{j i}^{(X\ell)}$ was 
calculated without any references to the external lines of unstable states. Instead, the amplitudes of associated scattering processes in which unstable heavy neutrinos appear only as \emph{internal} lines were studied; the resulting prescription for $\vep_{j i}^{(X\ell)}$ can be summarized as follows \cite{PilafUnd}. Consider the interaction \refer{Eq:N-to-h} between a hermitian scalar field $h$, heavy neutrinos $N_j$  (described in terms of Majorana fields $\psi_{N}^{j}$),   and (approximately) massless SM (anti)neutrinos, described in terms of Majorana fields $\psi_{\nu}^{i}$. Suppose that the matrix of propagators of heavy neutrino Majorana fields $\psi_{N}^{j}$ has 
the following form ($\cC$ is the charge-conjugation matrix)~\footnote{With apologies to 
  the reader for the proliferation of different fonts; unlike the tree level masses pole
  masses $\tm$ are in general {\em complex}.}
\eq{\label{Eq:D-as}
\hat{\sG}(p)
=
i\,\hat{\zeta}\,
[p^2-\tm^2]^{-1}[\ds{p}+\tm]\,
\hat{\zeta}^{\,{\rmt}}\,\mathcal{C}^{-1}
+\text{[non-pole part]}
\,,
}
where the matrix of pole masses 
\eq{\label{Eq:m-matr-Maj}
\tm=\diag(\tm_1\,,\  \tm_2\,,\ \tm_3)\,,
}
is diagonal with positive real parts ${\rm Re}(\tm_a) > 0$, while its imaginary part 
gives the total decay widths. The residue matrices $\hat\zeta$
can be written as
\eqs{\label{Eq:Zeta-hat}
\hat{\zeta}\,&=&\zeta_L\otimes P_L\, + \,\zeta_R \otimes P_R\,,\nn\\
\hat{\zeta}^{\,\rmt}&=&\zeta_L^{\ \rmt} \otimes P_L 
                                          + \zeta_R^{\ \rmt} \otimes P_R\,,
}
with 3$\times$3  matrices $\zeta_{L,R}$ carrying only family indices, and chiral projections $P_{L,R}$; clearly, at tree level, in the basis of mass eigenstates one has $\ze_L=\ze_R=\id$. 
If these matrices are known, 
the CP-asymmetry \refer{Eq:ep-def-nowa} can then be calculated with 
the aid of the following formula \cite{PilafUnd}
\koment{PDCN20,PDCN17,PDCN12}
\eq{\label{Eq:ep-form}
\vep_{j i}^{(h\nu)}
=
\frac{|\cY^{R}_{j {i}}|^2-|\cY^{L}_{j {i}}|^2}
     {|\cY^{R}_{j {i}}|^2+|\cY^{L}_{j {i}}|^2}
\,,
}
with \koment{str.PDNC26}
\eqs{
\label{Eq:cY-L}
\cY^{L}_{j {i}}
&=& \cY^{(h)}_{k {i}}(\zeta_L)^{k}_{\ j}+\ldots\,,\\
\label{Eq:cY-R}
\cY^{R}_{j {i}}
&=& \cY^{(h)\star}_{k {i}}(\zeta_R)^{k}_{\ j}+\ldots\,,
}
where the ellipses indicate contributions of corrections to external lines of $h$ and $\psi_{\nu}$ fields, as well as loop corrections to the 1PI vertices (which are negligible in $\TeV$-scale leptogenesis). If heavy neutrinos were stable, the matrix $\zeta_R$ would be the 
complex conjugate of $\zeta_L$. In that case Eqs. \refer{Eq:cY-L}-\refer{Eq:cY-R}
are nothing more than the ordinary LSZ-reduction rules for calculating 
the S-matrix elements, see e.g. \cite{Becchi}. Similarly, the CP-asymmetry $\vep_{j i}^{(Z\nu)}$ 
can be calculated with the aid of Eq. \refer{Eq:ep-form}, with the following replacements 
\eqs{
\nn 
\cY^{L}_{j i}
&=& \cF^{(Z)}_{k i}(\zeta_R)^{k}_{\ j}+\ldots\,,\\
\nn 
\cY^{R}_{j i}
&=& \cF^{(Z)\star}_{k i}(\zeta_L)^{k}_{\ j}+\ldots\,,
}
(the change of chirality is caused by $\gamma^\mu$). The enhancement effect
that underlies resonant leptogenesis is due to the $\ze_{L,R}$ matrices which contain the factors $\sim(\tm^2_1 - \tm_2^2)^{-1}$ etc. (see below).

{\begin{figure}[t]
\hspace{0.0cm}
\centering
\includegraphics[scale=1.25]{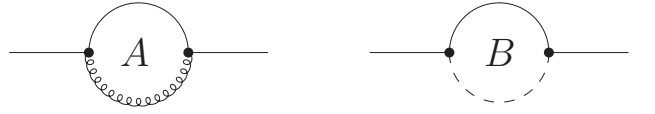}
\caption{One-loop contributions to $\sM_{L,R}$ and $\sZ_{L,R}$ in the Landau gauge. In CSM external lines represent heavy neutrino Majorana fields. Solid internal lines represent 
all fermions (with the exception of quarks), with dashed/springy
lines denoting scalars/massive vectors.}
\label{Diag:SelfEn}
\end{figure}

To find the $\ze_{L,R}$ matrices we use the prescription given in \cite{AL}, to which we
also refer for further details. Adopting some renormalization scheme, 
let $\widetilde{\Ga}(-p,p)$ be the matrix of renormalized 1PI two-point functions 
(inverse propagators) of the Majorana fields $\psi_{N}^{j}$  
\eqs{\label{Eq:Gamma2} 
\widetilde{\Ga}(-p,p)
&=&
\cC\Big\{
\phantom{+}\!\!\!\!
\left(\ds{p}\sZ_L(p^2)-\sM_L(p^2)\right)P_L
+\qquad\nn\\
&{}&\,\,\,\,\,+\!
\left(\ds{p}\sZ_R(p^2)-\sM_R(p^2)\right)P_R
\Big\}
\,\, ,
\qquad
}
where matrices $\sM_{L,R}$ and $\sZ_{L,R}=\mathds{1}+\cO(\hbar)$ carry only family 
indices. Now let $\bM^2_L(p^2)$ be the following matrix (with $s\equiv p^2$)
\eqs{\label{Eq:bM2-LR}
\bM^2_L(s)
&\equiv&
\sZ_L(s)^{-1}\,\sM_R(s)\,\sZ_R(s)^{-1}\,\sM_L(s)\,.
}
Then the propagator of $\psi_N^j$ has the form \refer{Eq:D-as} where the 
(complex) pole masses $\tm$ are solutions to 
\eq{\label{Eq:gapEq}
\det(s\mathds{1}-\bM^2_L(s))\bigg|_{s=\tm_{a}^2} = 0\,, 
}
while the columns of $\zeta_{L,R}$ matrices are given by vectors $\zeta_{L,R[a]}$ 
\eq{\nn
\zeta_X=
\bigg[
\Big[\zeta_{X[1]}\Big]\,\Big[\zeta_{X[2]}\Big]\,
\Big[\zeta_{X[3]}\Big]
\bigg]\,,
\qquad X=L,R,
}
which are obtained in the following way.  Let $\xi_{[a]}$ be an eigenvector of 
$\bM^2_L(\tm_{a}^2)$, with eigenvalue $\tm_{a}^2$  
\koment{str.Z17}
\eq{\label{Eq:xi-Eig}
\bM^2_L(\tm_{a}^2)\,  \xi_{[a]} 
= 
\tm_{a}^2\,
\xi_{[a]}\,,
} 
and obeying the following normalization condition 
\eq{\label{Eq:norm-cond}
\xi_{[a]}^{\,\, \rmt} \, 
\sM_{L}(\tm_{a}^2)\, 
\xi_{[a]}
=\tm_{a}\,, 
}
then \koment{str.Z30} 
\eq{\label{Eq:zeta-L-a}
\zeta_{L[a]}=\cN(a)\,\xi_{[a]}\,,
}
with a normalizing factor 
\eq{\label{Eq:cN-final}
\cN(a)
=
\Big\{
1-
\frac{1}{\tm_{a}}
\xi_{[a]}^{\ \rmt}\,\sM_L(\tm_{a}^2)\,
\bM^2_{L}{}^{\prime}(\tm_{a}^2)\,\xi_{[a]}
\Big\}^{-1/2}
\,,
}
where 
$
\bM^2_L{}^{\prime}(s) 
\equiv 
\dd\bM^2_L(s)/\dd s\,,
$
and
\koment{str.Z16}
\eq{\label{Eq:zeta-R-a}
\zeta_{R[a]}
=
\frac{1}{\tm_{a}}
\sZ_R(\tm_{a}^2)^{-1}
\sM_L(\tm_{a}^2)\,
\zeta_{L[a]}\,.
}

For heavy neutrinos $\tm_{a}\neq 0$ and the corresponding eigenspaces are one-dimensional. Thus the above prescription is all we need to calculate the CP-asymmetries $\vep_{ji}$ (for a generalization to massless or Dirac fermions, as well as the discussion of reality properties of $\ze_{L,R}$ matrices, see \cite{AL}). To obtain the required numerical values of $\sZ_{L,R}$ and $\sM_{L,R}$ matrices, we use the one-loop formulae given in \cite{AL}; these are 
valid for a general renormalizable model in the Landau gauge and correspond to 
the diagrams shown in Fig. \ref{Diag:SelfEn}.
Since $\zeta_{L[a]}$ is an eigenvector, the ordinary quantum-mechanical perturbation theory for discrete spectra (more precisely, its generalization to non-hermitian matrices) indicates that the components of $\ze_L$ are enhanced whenever masses of fermions are approximately degenerate. This in turn causes the enhancement of the CP-asymmetry \refer{Eq:ep-form}, and thus lepton asymmetry, dubbed ``resonant leptogenesis" \cite{PilafRL1}.

Some remarks are in order. While the above prescription for finding $\ze_{L,R}$ matrices 
is, in principle, independent of the choice of basis in the space of fields, we apply it in the 
basis of tree level mass eigenstates in which loop calculations are done. In this basis the $\ze_{L,R}$ matrices (unlike the $V_0$ matrix that diagonalizes the tree-level mass matrix itself, cf. Eq. \refer{Eq:DiagMas}) are numerically small perturbations of the identity matrix, for all cases studied below. 
It is also worth stressing that we completely neglect the masses of light neutrinos circulating 
in loops as in Fig. \ref{Diag:SelfEn}; this is justified since contributions of these masses are subdominant in $Y^\nu$, as can be easily checked from the mentioned generic one-loop 
formulae. In light of this fact, our choice $U_0=\id$ (rather than  $U_0=U_{MNS}$) in Eq. \refer{Eq:DiagMas} for light neutrino states is self-consistent.   

\begin{figure}[t]
\hspace{0.0cm}
\centering
\includegraphics[scale=1.8]{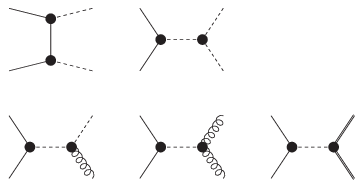}
\caption{Annihilation diagrams $N_iN_i\to XY$ induced by $y_M$. Standard Model fermions are depicted as double solid lines.}
\label{Diag:NNanni}
\end{figure}

\subsection{Boltzmann equations}

To determine the lepton number asymmetry one has to solve the Boltzmann equations 
(BEQs) in the context of an expanding universe  \cite{Kolb-Turner-Book}. For the  
CSM the full set of equations would be close to unmanageable due to the large number 
of degrees of freedom and possible processes involved, and one therefore has to resort 
to several simplifying assumptions. A first such assumption is that the elastic processes 
are fast, so that all species are in kinetic equilibrium, having the occupancy given by 
the Fermi-Dirac/Bose-Einstein distributions
\kE\eq{\label{Eq:f=}
f({\bf p})=\{\exp[(E({\bf p})-\mu)/T]\pm 1\}^{-1}\,.
} 
Secondly, in order to reduce the large number of independent distribution functions 
$f$ (or, equivalently, the associated chemical potentials $\mu$), we assume that 
all the interactions described by the Lagrangian density \refer{Eq:Lagr-tot}, with the 
exception of those triggered by $Y^\nu$ or $Y^M$, are in chemical equilibrium. %
\kF\footnote{Recall that, when the reaction $i + j \to k + l$
between particles $i,\ j,\ k $ and $l$ is in chemical equilibrium, then the corresponding
chemical potentials obey the relation $\mu_i + \mu_j = \mu_k + \mu_l$.
}
Note that in TeV scale leptogenesis this assumption is justified for the SM Yukawa 
couplings \cite{FMinL}.  A further simplification is achieved by assuming 
that the non-perturbative SM interactions that violate $B$ and $L'_i$ symmetries 
down to the combinations $B- L'$ and $L'_i-L'_j$ are also in equilibrium; direct analysis of these processes \cite{SFAL,PilafSMM} indicates that this assumption is  reasonable for $ T\gtrsim 80 \GeV$. Note that, although $X\equiv(B-L')$ unlike $(B-L)$ or $(L'_i-L'_j)$ has a $X$-$X$-$X$ anomaly, it does not have anomalies in the presence of the SM gauge field background, and thus it is preserved by sphalerons. In other words, 
$B-L'$ and $L'_i - L'_j$ are violated only by $Y^\nu$ induced interactions, and only these interactions contribute to the Boltzmann equations for the densities of these {\em differences}, see Eqs. (\ref{Eq:BoEqL}) and (\ref{Eq:BoEqBmL}), i.e. spaleronic interactions cancel out (clearly, these equations must then still be supplemented
by the ones for the heavy neutrino densities, see (\ref{Eq:BoEqN})).

Under these circumstances, there are four independent chemical potentials for the SM species, 
which correspond to these global symmetries, and, in addition, after the 
electroweak phase transition, to the electric charge; however, the electric neutrality of the universe allows us to express the charge potential as a linear combinations of the remaining ones 
\cite{Harvey-Turner}. For our purposes, it is convenient to choose the light 
neutrinos' potentials as independent ones
\kC\kG\eq{
\mu_{\nu_i}\equiv\mu_{e_i}+\mu_{W_+}\,.
}
Neglecting masses of SM particles, $\mu_{\nu_i}$ can be expressed in terms of individual 
SM lepton number densities $n_{L'_i}$ in the broken phase of the SM,
which simplifies to 
\footnote{This result can be easily obtained by repeating the analysis of \cite{Harvey-Turner}  \emph{without} the assumption that $\mu_{\nu_i}\equiv \mu_{\nu}$ for all flavors $i$. Note that in the present context there are no rapid flavor-mixing interactions; in particular matrix elements of $Y^\nu$ are small.}
\eq{\label{Eq:mu-nu-i-div-by-T}
\frac{\mu_{\nu_i}}{T} =
 \frac{166\, n_{L'_i}+16 (n_{L'_j}+n_{L'_k})}{75\, T^3}\,,
}  
where $i\neq j\neq k \neq i$. Similarly, the lepton-number density can be expressed 
in terms of the $B-L'$ density as follows \cite{Harvey-Turner} 
\eq{\label{Eq:Lep-Bar-rel}
\sum_i n_{L'_i} = -\frac{25}{37} 
\Big[n_B-\sum_i n_{L'_i}\Big] \,.
}
To simplify the BEQs for the number densities, we approximate the occupancies 
\refer{Eq:f=} with the Maxwell-Boltzmann distributions
\eq{\label{Eq:f=MaxBo}
f\approx\exp[-(E-\mu)/T]\,.
}
This allows to perform some of the momentum integrals analytically 
(and should not lead to errors bigger than $20\%$ \cite{PilafUnd}).    
In this approximation, the distribution of heavy neutrinos $N_i$ can be written as 
\eq{
f_{N_i} = \exp(-E/T)\, \frac{n_{N_i}}{n_{N_i}^{\rm EQ}}\,,
}
where $n_{N_i}$ is the number density of ${N_i}$, while 
$n_{N_i}^{\rm EQ}$ is the value $n_{N_i}$ in the chemical equilibrium  
(i.e. the one corresponding to the vanishing chemical potential, as indicated by various $y_M$-induced annihilation processes, e.g. $N_i N_i \to t \bar t\,$, see Fig. \ref{Diag:NNanni})  that reads (see e.g. \cite{ThermalLepto})
\kK\eq{
n_{N_i}^{\rm EQ}
=\frac{m_N^2\,T}{\pi^2}K_2(m_N/T)\,,
}   
with $K_\ell(z)$ denoting the modified Bessel functions of the second kind (similarly, thermally averaged decay widths lead to the appearance of $K_1(z)$ in BEQs below) and $m_N=y_M v_\phi/\sqrt{2}$ being the Majorana mass (the mass splitting due to $Y^\nu$ is negligible as far as the distributions of heavy neutrinos are concerned). Since the Majorana mass in the present model 
originates from the vacuum expectation value of $\phi$, we assume below that the baryon asymmetry is produced after spontaneous breaking of $B-L$, from initially symmetric state. The analysis of phase transition will be given elsewhere.

Due to the expansion of the universe it is convenient to write the BEQs for 
densities normalized to the entropy density $s(T)\propto T^3$ 
(see e.g. \cite{Kolb-Turner-Book})
\eq{   \label{Yscr}
{\mathscr{Y}}_X=\frac{n_X}{s(T)}\,,
}
as functions of the following `time' variable 
\eq{
 z=z(T)=\frac{m_N}{T}\,.
}

With these approximations, it is fairly easy to write BEQs for the densities of (approximately)
conserved charges. In particular, the symmetries $B-\sum_i L'_i $ and $L'_i-L'_j$ are violated 
only by the $Y^\nu$-induced interactions (but not by anomaly induced instanton
processes). Denoting by $D^{L'_i}$ the appropriate combinations of  averaged 
squared-amplitudes of $Y^\nu$-induced processes that 
violate $L'_i$, one can write the relevant BEQs for the 
densities of these non-anomalous charges in the following form
\kA\eq{\label{Eq:BoEqL}
s(T)H(T)z\deru{}{z}\left[{\sY}_{L'_i}-{\sY}_{L'_j}\right]
=D^{L'_i}-D^{L'_j}\,,
}
\kI\eq{\label{Eq:BoEqBmL}
s(T)H(T)z\deru{}{z}\big[{\sY}_{B}-\sum_i {\sY}_{L'_i}\big]
=-\sum_i D^{L'_i}\,,
}
where $H(T)\propto T^2$ is the expansion rate of the universe \cite{Kolb-Turner-Book}. 
The BEQs for individual ${\sY}_{L'_i}$ follow then immediately from 
(\ref{Eq:Lep-Bar-rel}). 
The dominant contributions to $D_i^{L'}$ come from decays and inverse decays of 
heavy neutrinos (as well as the subtraction of their real intermediate states from the 
associated scattering processes, the latter has been taken care of by following the approach of \cite{PilafUnd}; in particular decays of heavy neutrinos with equilibrium distributions ${\sY}_{N_i}={\sY}_{N_i}^{\rm EQ}$ do not contribute to $D^{L'_i}$ given below, in agreement with the Sakharov conditions \cite{Sakh}%
\kJ). 
To the first order in small parameter \refer{Eq:mu-nu-i-div-by-T}, $D^{L'_i}$ have the 
form (for a discussion of thermally averaged rates, see e.g. \cite{ThermalLepto})
\kB\eq{\nn
D^{L'_i}
=
\frac{m_N^3}{\pi^2\,z}K_1(z)
\sum_j
\Big\{
\Big[\frac{{\sY}_{N_j}}{{\sY}_{N_j}^{\rm EQ}}-1\Big]\De_{j i}
-
\frac{\mu_{\nu_i}}{T}\Si_{j i}
\Big\}\,,
}
with
\kB\eq{\label{Eq:Si}
\Si_{j i}=\sum_{X,\ell}
\left[\Ga(N_j\to X \ell_i) + \Ga(N_j\to \bar{X} \bar{\ell}_i)\right]\,,
}
\eq{\label{Eq:De}
\De_{j i}=\sum_{X,\ell}
\left[\Ga(N_j\to X \ell_i) - \Ga(N_j\to \bar{X} \bar{\ell}_i)\right]\,,
}
where the summation runs over different decay channels with $\ell_i\in\{e_i,\,\nu_i\}$ denoting a charged or neutral lepton of $i$th flavor. Clearly, $\Si_{j i}$ can be calculated with a good accuracy at the tree level. In calculating $\De_{j i}$, the CP-asymmetries introduced in the previous section are crucial (cf. Eq. \refer{Eq:ep-def-nowa}) 
\eq{\label{Eq:De-new}
\De_{j i}=\sum_{X,\ell}
\vep^{(X\ell)}_{ji} \times
\Big[\Ga(N_j\to X \ell_i) + \Ga(N_j\to \bar{X} \bar{\ell}_i)\Big]\,.
}

For given ${\sY}_{N_j}={\sY}_{N_j}(z)$, Eqs. 
\refer{Eq:BoEqL}-\refer{Eq:BoEqBmL} form a system of three equation for 
three independent functions $n_{L'_i}$, cf. Eqs. \refer{Eq:mu-nu-i-div-by-T} 
and \refer{Eq:Lep-Bar-rel}. They have to be supplemented with 
three more equations for ${\sY}_{N_j}$ 
\kD\eq{\label{Eq:BoEqN}
s(T)H(T)z\deru{{\sY}_{N_j}}{z}
=D^N_j+A^N_j+S^N_j\,,
}
where $D^N_j$ represents the effects of $Y^\nu$-induced decays of a heavy 
neutrino $N_j$ 
\kL\eq{
D^N_j=
-\Ga_{N_j} 
\frac{m_N^3}{\pi^2\,z }K_1(z)
\left[\frac{{\sY}_{N_j}}{{\sY}_{N_j}^{\rm EQ}}-1\right]
\,,
}
where  the total decay width  is determined via
$\Ga_{N_j} \approx 
- {\rm Im } (\tm_j/2)$ with the pole mass obtained from Eq. \refer{Eq:gapEq}. 
While this is the standard contribution to BEQs that occurs also
in `minimal' leptogenesis scenarios, the other two contributions labeled $A_j^N$ and $S_j^N$
are absent in such a minimal framework, as they 
represent contributions arising from the new scalar field $\phi$. More specifically,
$A^N_j$ describes the rates of $y_M$-induced annihilation processes of 
heavy neutrinos (see Fig. \ref{Diag:NNanni}), and $S^N_j$ represents the 
rates of inelastic scatterings shown in  \mbox{Fig. \ref{Diag:NNscatt}}. 
\koment{str.ROZP51}
\eq{\nn
A^N_j=
-\frac{m_N}{64\pi^4\,z}\cK\big[\si_j\big]\Bigg\{\bigg[\frac{{\sY}_{N_j}}{{\sY}_{N_j}^{\rm EQ}}\bigg]^2-1\Bigg\}\,,
}
\eq{\nn
S^N_j=
-\frac{m_N}{64\pi^4\,z}\sum_{i\neq j}\cK\big[\si_{j\to i}\big]
\Bigg\{
\bigg[\frac{{\sY}_{N_j}}{{\sY}_{N_j}^{\rm EQ}}\bigg]^2
-
\bigg[\frac{{\sY}_{N_i}}{{\sY}_{N_i}^{\rm EQ}}\bigg]^2
\Bigg\}\,,
}
where 
\eq{\label{cK-int}
\cK\big[\si\big]
=2\int\limits^\infty_{(2m_N)^2}
{\dd s}\,\sqrt{s}\,(s-4 m_N^2)\,K_1(\sqrt{s}/T)\,\si(s)\,.
}
Here $\si_j(s)$ denotes the total cross-section for the processes ${N_jN_j\to XY}$ shown in Fig. \ref{Diag:NNanni}, while $\si_{j\to i}(s)$ is the cross-section for the processes ${N_jN_j\to N_iN_i}$ (Fig. \ref{Diag:NNscatt}). (Note that these cross-sections are summed, rather than averaged, over initial spin states, and that the form of the lower limit in \refer{cK-int} appears because heavy neutrinos turn out to be the heaviest particles in the model; for a discussion of thermally averaged cross-sections see e.g. \cite{ThermalLepto}). 
Despite the huge hierarchy between $y_M$ and $Y^\nu$, both classes of processes are equally important, since $D^N_j$ and $A^N_j$ ($S^N_j$) have different dependencies on $z$. The importance of $y_M$-induced processes was emphasized in \cite{Anni}  (see also \cite{BLJ,BLNJ} for a discussion in the context of local $B-L$ models). Their presence is a main difference between models with spontaneous lepton-number violation and the ``minimal leptogenesis", in which $Y^\nu$ is the sole source of non-conservation of both, $B-L'$ as well as the number of heavy neutrinos. 
In particular, they keep heavy neutrinos in thermal equilibrium at early times (see also the discussion in the next subsection). 

\begin{figure}[t]
\hspace{0.0cm}
\centering
\includegraphics[scale=1.8]{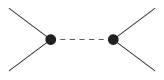}
\caption{$\cO(y_M^2)$ inelastic processes $N_iN_i\to N_jN_j$.}
\label{Diag:NNscatt}
\end{figure}}

\subsection{Results}

For  $Y^\nu = \sqrt{2} m_D/v_H$ we use the Casas-Ibarra parametrization \refer{Eq:CaIbPar}, assuming inverted ordering ($m_{\nu3}<m_{\nu1}<m_{\nu2}$~\footnote{The inverted ordering 
 of light neutrino masses is just a choice that we made for the scan over remaining 
 parameters, but not necessarily a prediction of our model.}) 
with central values of all neutrino oscillation parameters, including the Dirac phase 
of the matrix $U_{MNS}$, given in \cite[Table 14.7 on page 252]{PDG2014}. For both  
unconstrained Majorana phases in $U_{MNS}$ we take the value $2\pi/5$, 
while for the lightest neutrino we assume $m_{\nu3}=1.08\times 10^{-3}\,\eV$. 
As to the complex angles of Casas-Ibarra matrix, a set of values 
(in the standard CKM-like parametrization) that works is the following
\eq{
\al=\frac{9 \pi }{25}+\frac{33 i}{25}\,,
\quad
\be=\frac{6 \pi}{5}+\frac{18 i}{25}\,,
\quad
\ga=\frac{4 \pi }{5}+\frac{11 i}{25}\,,
}
so that
\eqs{
\RCI&\approx&
\nn
\left(
\begin{array}{ccc}
 -0.15+2.0 i & -2.2+0.1 i & -0.75-0.63 i \\
 2.2+0.13 i & 0.02+1.9 i & -0.49+0.68 i \\
 -0.18-0.06 i & -0.31-0.42 i & 1.0-0.13 i \\
\end{array}
\right)\,.
}
While the matrix elements of $\RCI$ are of order $\cO(1)$, the above form of $\RCI$ ensures that 
the decay width of one of the heavy neutrinos is suppressed in comparison with the other two. This allows for a sufficient departure from equilibrium in the range of temperatures in which $B$-violating processes are still fast. Let us emphasize that there is nothing unique about 
this choice of parameters, which we have adopted here simply because it does give 
the right order of magnitude for the lepton asymmetry; there may thus exist other
viable ranges of parameters.

For the integration of the  BEQs, we assume that for $T = 10\, m_N$  heavy neutrinos 
are in equilibrium ($\sY_{N_i}=\sY_{N_i}^{\rm EQ}$) while all leptonic asymmetries $\sY_{L'_i}$ vanish. The resulting baryon asymmetry $\sY_{B}$ (for $T=100 \GeV$, when 
$B$-violating interactions decouple) is given in Table \ref{Tab:Points}. It corresponds 
directly to the present value $\sY_{B0}$ predicted by CMS, under assumption 
of entropy conservation. Using the present entropy to photon ratio 
$s\approx 7 n_\ga$ \cite{Kolb-Turner-Book}, 
the baryon-to-photon ratio \refer{Eq:nB} translates into 
\eq{  \label{YB0}
\sY_{B0}\approx 8.6 \times 10^{-11}\,,
}
we thus see that the values in Table \ref{Tab:Points} agree quite well with the data. 
Although our input value for $s$ does not include the contribution form Majorons 
to the total entropy, their inclusion would not affect our results in any essential way.

\begin{figure}[t]
\centering
\includegraphics[width=0.45\textwidth]{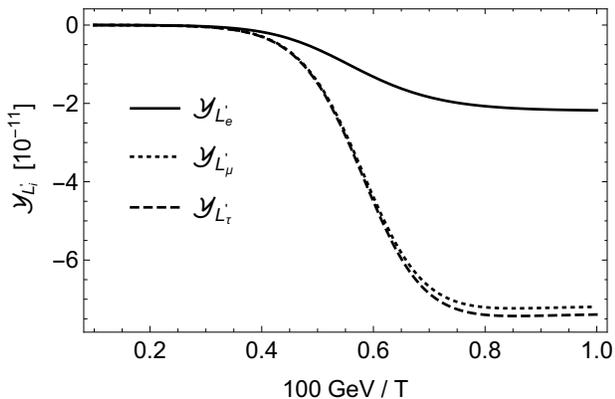}
\caption{Leptonic asymmetries $\sY_{L'_i}$ for different flavors as a function of inverse temperature for the first point in Table \ref{Tab:Points}.
}
\label{Plot:L}
\end{figure}

\begin{figure}[t]
\hspace{0.0cm}
\centering
\includegraphics[width=0.45\textwidth]{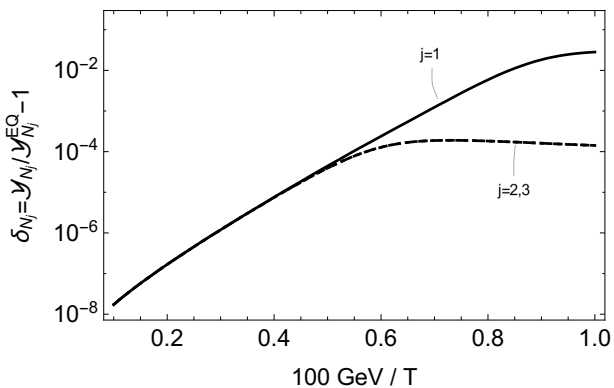}
\caption{
Departure $\de_{N_j}\equiv{{\sY}_{N_j}}/{{\sY}_{N_j}^{\rm EQ}}-1$ of different flavors of heavy neutrinos from thermal equilibrium  
 as a function of inverse temperature for the first point in Table \ref{Tab:Points}. Only one flavor has a significant departure in the range of temperatures in which $B$-violating processes are fast. 
}
\label{Plot:N}
\end{figure}

\begin{figure}[t]
\centering
\includegraphics[width=0.45\textwidth]{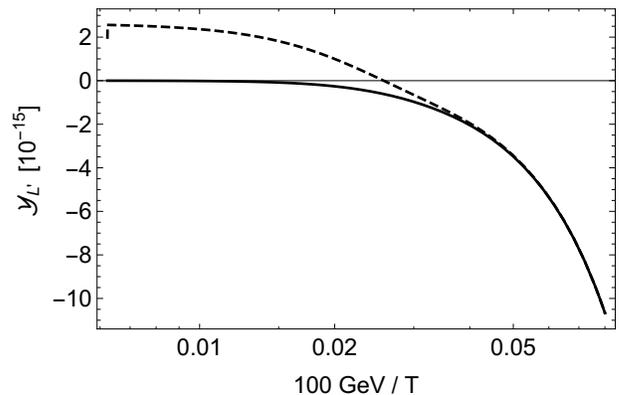}
\caption{{
Total leptonic asymmetry $\sY_{L'}$ for the first point in Table \ref{Tab:Points}. The solid line corresponds to thermal initial abundance of heavy neutrinos (for $T_0=10\,m_N$). The dashed line corresponds to vanishing initial abundance of heavy neutrinos (for $T_0=10\,m_N$). In both cases vanishing of initial asymmetries $\sY_{L'_i}$ is assumed.   
}}
\label{Plot:L-tot}
\end{figure}

The integral curves of Boltzmann equations are illustrated, for the first point in Table \ref{Tab:Points}, in Figs. \ref{Plot:N} and \ref{Plot:L}. Note that, due to fast $y_M$-induced interactions, heavy neutrinos depart from equilibrium for relatively small temperatures; 
this behavior was also observed in \cite{Anni}. Nonetheless, our analysis shows  
that successful \emph{resonant} leptogenesis is possible. 
We note that the $y_M$-induced processes justify our assumption about initial thermal abundance of heavy neutrinos. In fact, the present baryon asymmetry is essentially independent of the distribution of heavy neutrinos for $T\gg m_N$. This can be seen in Fig. \ref{Plot:L-tot}, where the dynamically generated lepton asymmetry for thermal initial distribution of heavy neutrinos (solid line) is compared with its counterpart for vanishing initial abundance (dashed line).  
We also stress that the effects of thermal corrections to particles' masses were 
neglected here, and will be discussed in a separate publication, where also issues 
related to the phase transition will be addressed.

Let us also mention that a similar analysis can be performed for the model with an 
extended scalar sector that was proposed in our previous work \cite{U3model}. The
result is that resonant leptogenesis does not work in that case, even though for
that model Eq. (\ref{Eq:YM-deg}), and thus the near degeneracy of heavy neutrino masses,
is an automatic consequence of spontaneous
symmetry breaking. The reason is that with this extended scalar sector, the minimization condition for the pseudo-Goldstone boson potential (Dashen's condition \cite{DASHEN,WeinT2}) requires $\RCI$ to be real, whence the unitary
matrix $U_{MNS}$ in \refer{Eq:CaIbPar} is the sole source of CP-violation. This CP-breaking 
turns out to be insufficient to overcome the rapid $y_M$-induced interactions that keep 
heavy neutrinos in thermal equilibrium: the processes of the type $A$ and $S$ above
are generically faster in the presence of more scalar fields.

\section{Dark matter}
We next turn attention to the Goldstone boson that accompanies the spontaneous breaking
of $(B-L)$ symmetry. We note that this particle comes `for free' with our model, and
provides a natural `habitat' for lepton number violation, a feature that we
exploited already in the previous section. However, spontaneous symmetry breaking is
not enough for a possible explanation of Dark Matter, because for that the Goldstone 
boson must acquire (an albeit tiny) mass by a separate mechanism

\subsection{Explicitly Breaking $(B-L)$ Symmetry} \label{Sec:Break-B-min-L}

As already eveident from the previous section an important feature of the 
CSM in its unbroken phase is the lepton number symmetry under which
also the new scalar $\phi$ transforms non-trivially. However, rather than focusing on this
symmetry separately, we will now consider the $(B-L)$ transformations which likewise 
leave the Lagrangian \refer{Eq:Lagr-tot} invariant
\bea \nn
(L^i_\al(x), \bar E^i_\da(x), \bar N^i_\da(x))  
     \; &\rightarrow& \;
 e^{- i\omega} (L^i_\al(x), \bar E^i_\da(x), \bar N^i_\da(x))    \nnn[2mm]
(Q^i_\al(x), \bar U^i_\da(x), \bar D_\da ^i(x))  
     \; &\rightarrow& \; 
e^{\frac13 i\omega}(Q^i_\al(x), \bar U^i_\da(x), \bar D_\da ^i(x))  
  \nnn[2mm]
\phi (x)\; &\rightarrow& \; e^{-2i\omega} \phi(x) \nn
\eea
The appearance of both barred and unbarred spinors here is dictated by
demanding invariance of the Yukawa interactions \refer{LY}; clearly, the resulting 
transformations of the Dirac fields \refer{Eq:Dirac-Field} and (\ref{Dirac2}) are indeed non-chiral. 
The reason for considering $(B-L)$ rather than just lepton number is that this symmetry is anomaly free (see e.g. \cite{WeinT2}), which ensures that after spontaneous symmetry breaking 
$\phi(x)$ contains, in addition to a real massive scalar, also a Goldstone boson  
that remains massless to all orders in perturbation theory thanks to the vanishing
$(B-L)$ anomaly. Some of these properties are better visible  in the 
exponential parametrization 
\beq\label{A}
\phi (x) = \frac{1}{\sqrt{2}}
\big( v_\phi + R(x)\big) \exp\big( 2\,i A(x)\big)
\eeq
where we split the complex field $\phi(x)$ into a modulus $R(x)+v_\phi$ and a phase $A(x)$.
The latter can be absorbed into a redefinition of the fermions
\eqs{\label{phidef}
\chi_\al(x) \; &\rightarrow& \;\; \chi_\al(x)^{\rm NEW} \equiv \exp\big[-i(b-\ell) A(x)\big] \chi_\al(x) 
\,,\nn\\
\bar\chi_\da(x) \; &\rightarrow& \;\; \bar\chi_\da(x)^{\rm NEW} \equiv \exp\big[+i(b-\ell) A(x)\big] \bar\chi_\da(x)
\,,\nn\\
}
where $\chi_\al(x)$ is any CSM Weyl field, and $(b-\ell)$ is its charge under U(1)$_{B-L}$.
Due to the exact $(B-L)$ invariance the field $A(x)$ then appears in the new Lagrangian 
with redefined fields only via derivative couplings originating from the kinetic terms 
of $\phi$ and $\chi_\al$'s, to wit, 
\beq
\cL_{int} \, \propto \, (b-\ell)\,  \bar {\chi}_\da\, \anti\sigma^{\mu\,\da\be}\,  {\chi}_\be\, \partial_\mu A
\eeq
(we drop the label NEW) and the kinetic term
\beq
\cL_{kin}^A = 2 \, v_\phi^2 \, \partial^\mu\! A \partial_\mu A  \, + \, \cdots
\eeq
which is not canonically normalized; 
the dots stand for couplings of $A(x)$ to the real scalar $R(x)$.
In this picture the fact that $A(x)$ couples only via derivatives is  completely manifest. 
\footnote{
Due to the small mass term to be introduced below, cf. (\ref{ma}), there will also
arise {\em non-derivative} effective couplings to SM fields which are very small \cite{Latosinski}.}

The above parametrization in terms of redefined fields will be referred to as the `exponential picture' (as opposed to the `linear picture' introduced in Sec. \ref{Sec:Lagrangian}). 
In particular $R(x)$ is the counterpart $\sqrt{2}{\rm Re}(\phi)$ from Sec. \ref{Sec:Lagrangian}, 
that is, it describes mainly the extra massive scalar boson $\cphi$ with a small admixture of the 
SM-like particle $h$, cf. Eq. \refer{eqn:Mixing}, while $A(x)$ corresponds to the 
Goldstone mode $a(x)$ up to normalization. The shift symmetry in the Goldstone 
field, $A(x)\to A(x)+\rm{const}$, is manifest in the exponential picture, but the price to pay
is that manifest renormalizability is lost. 

Although the field $A(x)$ thus cannot acquire a mass term within the framework of 
relativistic QFT in flat spacetime, we now recall a folklore theorem (still based 
on somewhat heuristic reasoning, cf.  \cite{Holman:1992us,Kallosh:1995hi,Bousso})
according to which there cannot exist exact {\em continuous} global symmetries in 
a quantum theory of gravity. This then leaves two options: either $(B-L)$ is gauged, 
in which case there is an extra massive $Z'$ boson, or otherwise the $(B-L)$ symmetry 
is broken explicitly by quantum gravity effects. 

 The former possibility has been studied both within a GUT context 
(in which case $Z'$ would be very heavy) or in a `low energy' realization with a $Z'$ boson 
whose mass is $\propto v_\phi^2$; a possible realization of the latter scenario within 
the CW context was investigated in detail in \cite{CW5a}.
Although we will not further consider this possibility here, let us note that for the CSM, 
gauging $(B-L)$ would give a very definite prediction for the mass 
of the $Z'$ vector boson. From the gauged kinetic term for $\phi$ 
\beq
\big(\pa_\mu\phi+2\ri q_{BL}Z'_\mu\phi\big)^\dagger
\big(\pa_\mu\phi+2\ri q_{BL}Z'_\mu\phi\big)
\eeq
we would get (after spontaneous symmetry breaking)
\beq
m_{Z'}=2\sqrt{2}q_{BL}\langle\phi\rangle=2q_{BL}v_\phi\,,
\eeq
The potential discovery of $\phi$ and knowledge of $v_\phi$ and $q_{BL}$ would thus
severely constrain the possible range of mass values for $Z'$, such that existing lower bounds 
on the mass of $Z'$ (that now exceed 4 TeV \cite{Exotics.png}) could already exclude
this possibility.

Because there is so far no evidence for a low lying $Z'$ vector boson, and because 
we wish to exploit the presence of the Goldstone boson in a different way by exploring 
its possible role as a Dark Matter candidate, we will here consider the 
second option, invoking (as yet unknown) quantum gravity effects, possibly in the form of a 
non-perturbative self-regularization of IR divergences, to generate a mass for the 
Goldstone boson. 
The non-perturbative breaking of $(B-L)$ symmetry via quantum gravity
was already considered in \cite{ABMS} which also invokes a gravity induced
mass for the Majoron to derive limits on its mass from the requirement that it should
not lead to over-closure of the universe. Although that work invokes a dimension 5
operator rather than a dimension 6 operator, as we do here, and does not appear 
to consider possible connections with Dark Matter, we note that,
interestingly, it also arrives at the conclusion that the scale of $(B-L)$ symmetry
breaking must not exceed $\cO(10\, \TeV)$. See also \cite{RBS} for a proposal along
these lines with gauged U(1)$_{B-L}$ and an extra scalar field as the Dark Matter candidate.

To implement the explicit symmetry breaking, we thus postulate 
the mass term 
\beq\label{amass}
\cL_A = \frac{v^4}{\MPL^2} \phi^2 \, + \, {\rm h.c.}
\eeq
which breaks $U(1)_{B-L}$ symmetry {\em explicitly} to its discrete subgroup
$\mathbb{Z}_2$. Unlike continuous symmetries, {\em discrete} symmetries are 
generally believed to be compatible with quantum gravity, which is our reason
for excluding dimension 5 operators, as the $\mathbb{Z}_2$ symmetry of the
CSM is thus preserved. Here $v$ is assumed to be of the same 
order of magnitude as $v_\phi$,  and the above mass term 
should thus be treated on a par with the tree-level Lagrangian. The inverse factor
of $\MPL^{-2}$ in (\ref{amass}) is included because this term is expected to be 
the low energy effective operator originating from quantum gravity.
Importantly,  (\ref{amass}) breaks $(B-L)$ symmetry only softly, and thus does not entail
new quadratic divergences, nor $(B-L)$ breaking dimensionless couplings, in 
analogy with the soft  terms in MSSM-like models.

Without spontaneous symmetry breaking the above mass term is completely negligible.
When $(B-L)$ symmetry is spontaneously broken, however, this term will manifest
itself in the form of a violation of the Goldstone Theorem, by endowing the Goldstone 
boson with a tiny mass and, in fact, a periodic potential for the Goldstone
field $A(x)$. The Goldstone theorem 
is a well known result of flat space QFT, but there is no reason to expect it 
to hold in the presence of a curved spacetime background or in the context 
of quantum gravity, and this is a possibility we wish to exploit here. A possible breakdown of the Goldstone Theorem 
in the framework of curved space QFT has already been discussed in the literature, 
see e.g. \cite{Goldstone1,Goldstone2,Goldstone3} all of which reach the conclusion that
in a curved background such as de Sitter space the Goldstone boson acquires a (small) mass.
Choosing $v \sim 1\,\TeV$ in formula (\ref{amass}) as an example we get
\beq\label{ma}
m_A  =\frac{{2}v^2}{\MPL}\,\sim\,  10^{-3} \eV\,.
\eeq
With the assumed small  quantum gravity induced mass and because of its very small
couplings to SM particles, we name the associated pseudo-Goldstone particle `minoron'. 

Importantly, the operator Eq. \refer{amass} in the exponential picture not only 
generates a mass term for the minoron, but also induces very small 
(and calculable)  \emph{non-derivative} couplings for the scalar field $A(x)$.
In particular the continuous shift symmetry $A(x) \rightarrow A(x) + const.$ is now 
reduced to a symmetry under discrete shifts, which implies that the induced potential for $A(x)$ must be a periodic function.

\subsection{Minorons as Dark Matter Candidates}
 
The mass estimate (\ref{ma}) lies very well within the range of mass values generally
accepted (or even desired) for Dark Matter constituents. Of course, in any such model
we have to ensure that the Dark Matter candidate cannot decay early on in the 
history of the universe, and therefore we assume that $m_A <2 m_\nu$. \footnote{For a discussion of $m_A >2 m_\nu$ case in the context of Dark Matter, see \cite{h1,h2,h3}.} 
In this section we briefly discuss the potential prospects for the minoron to be  a viable
Dark Matter candidate. In addition to its stability to decays, this requires that minorons 
must be created in sufficient amounts and in such a fashion that they can clump 
(as opposed to being thermally distributed like the CMB). There are obviously 
many analogies between the present proposal and axionic Dark Matter 
scenarios \cite{Sik,Alonso}, as the axion is also a 
pseudo-Goldstone particle. On the other hand, there are also differences -- in particular, there 
is no immediate link between the minoron and the strong interactions, unlike for the usual axion,
although axion-like couplings can be generated via higher loop corrections \cite{Latosinski}.

A main feature of any Dark Matter model concerns the possible interactions with
SM matter which must be small.
The coupling between the minoron and photons is of the loop origin. After summation over the helicity states of final photons, the amplitude for the processes $a\to \gamma \gamma $ can be 
bounded above by the following estimate
\eq{\label{Eq:cM-a-ga-ga}
\widetilde{\cM}
=
\Big\{\sum_{\rm spin} |\cM|^2 \Big\}^{1/2}
\lesssim \frac{1}{F} \frac{y_M\,e^2}{(4\pi)^2}\, p^2\,,
}
where $F$ is at least of the order of masses of particles circulating in the loops. Clearly
gauge-invariance of the $a \gamma \gamma$ vertex requires at least one 
momentum for each photon;  when the minoron is on-shell we have  $p^2 \sim m_A^2$. 
Taking $F=100\,\GeV$ and  $m_A=10^{-3}\eV$ we get
\eq{\label{Eq:Ga-a-ga-ga}
\Gamma( a\to \gamma \gamma)
=\frac{\widetilde{\cM}^2 }{16\,\pi\,m_A} \lesssim 10^{-48}\, \GeV\,.
}
Comparing this with the age of the universe 
($H_0 \sim  10^{-42}\, \GeV$) we see that the minorons can 
easily survive to the present epoch.  Nonetheless we should note that 
the decay width in Eq. \refer{Eq:Ga-a-ga-ga} is, in fact, overestimated by 
many orders of magnitude. First, the amplitude  \refer{Eq:cM-a-ga-ga} originates from 
multi-loop diagrams of the type discussed in \cite{Latosinski}, 
while in \refer{Eq:cM-a-ga-ga} we have included only coupling from vertices to which external lines are attached, as well as a single loop-suppression factor. Second, Goldstone bosons of \emph{non-anomalous} symmetries have derivative couplings to \emph{gauge-invariant} operators (see e.g. \cite{WeinT2}), thus additional powers of $p^2/v_\phi^2 \sim m_A^2/v_\phi^2$ should appear   
on the right-hand-side of \refer{Eq:cM-a-ga-ga}; 
while the minoron is pseudo-Goldstone boson, the explicit breaking of $(B-L)$ 
would itself introduce an additional factor $m_A^2/v_\phi^2$.

The minoron abundance is more difficult to estimate, and we can offer only some 
preliminary heuristic arguments at this point. The contribution to the density can come 
from three sources: particles, strings and domain walls.  Minorons, being lighter than light neutrinos, can decay only into photons but their lifetime is longer than the age of the Universe, so they are 
effectively stable -- therefore they pose no problem for the galaxy formation, nor for the nucleosynthesis. 
At the present time the relic thermal density of minorons is negligible. The minoron potential becomes
relevant when the field $\phi$ acquires its vacuum expectation value, and the minoron field 
decouples from other fields (its interaction with neutrinos is too weak to maintain equilibrium). 
The field starts to be dynamical when $3H\sim m_A$; $\langle\phi\rangle$ thus starts 
to differ from zero at $T\sim 1$ TeV. In this case both quantities are of  the order of $10^{-3}$ eV. 
The initial density of coherent oscillations is $\rho_{osc}\sim m_A^2v^2$ and after dilution it gives  a negligible contribution now. So we are left with strings (that decay very fast) and domain walls as 
the most important possible source of Dark Matter in the late history of the Universe 
relevant for the present day \cite{ZKO,Sikivie,Axion, RBS}.

We can write the approximate Lagrangian for $A(x)$ as
\beq
\cL_{minoron} = 2v^2_\phi g^{\mu\nu}\pa_\mu A\pa_\nu A-\frac14  v^2_\phi m_A^2 \big[1-\cos (4A)\big]
\label{minlagr}
\eeq
 Since $2A$ is a phase (see (\ref{A})) the 
period equals $\pi$ 
 (and not $2\pi$), and this is important for the stability of domain walls. For axion Dark Matter
 scenarios this stability is usually a problem, as it could lead to an over-closure of the Universe,
 but it is {\em not} a problem for the present scheme because the domain walls start to have a significant effect only in the present era, when the cosmological constant starts to dominate the evolution of the Universe.
 
We now assume that the domain wall connects two consecutive minima of the potential
(along the $z$-direction), for example $0$ and $\pi$. Neglecting time derivatives 
we have to solve the equation (where the prime denotes derivative with respect 
to the physical coordinate $z$)
\beq
A''(z) -\frac{m_A^2}{4}\sin (4A(z))=0
\eeq
with $A(-\infty)=0$ and $A(\infty)=\pi/2$. The solution reads
\beq
A(z) =\arctan\left(\re^{m_A z}\right)
\eeq
We can calculate the surface energy of the domain wall by
\beq
\si=\int_{-\infty}^\infty \rd z\ 2v_\phi^2\left(\left(\frac{\rd A}{\rd z}\right)^2+\frac{m_A^2}{8}(1-\cos4A)\right)
\eeq
with the result
\beq
\si=2m_A v_\phi^2
\eeq
Assuming $m_A\sim 10^{-3}$ eV and $v_\phi\sim 2$ TeV we get $\si\sim 2\cdot 10^{35}$ eV/m$^2$.
Assuming that these domain walls are very large, and that there is one wall per Hubble 
volume, the energy density thus comes out to be
\beq
\rho(t)\sim \si H(t)\sim \frac{t_0}{t}\ (\GeV/{\rm m}^3)
\eeq
where $t_0\sim 4\cdot 10^{17}$s is our present time. Remarkably, we thus arrive at the 
right order of magnitude for the present density of Dark Matter
\beq
\rho_{DM}\approx 1 \GeV/{\rm m}^3
\eeq

The presence of one or several large domain walls at the time of last scattering could have observable impact on the CMB spectrum especially for low $\ell$ (quadrupole) so possibly the domain walls should start decaying into (cold) minorons before the last scattering. Then the above estimate should be slightly changed since the energy density of particles decreases faster than that of domain walls. However,
this topic requires further study for a more precise analysis.

One can also note that the  self-interaction of massive minorons via interactions with 
right-chiral neutrinos (via box diagrams {\em \`a la} Euler-Heisenberg) gives similar 
values as would be required  by the Steinhardt-Spergel analysis of Dark Matter 
in the Abell cluster \cite{SpStein}. The values for the cross section are in the region 
$\sigma\sim m_A\cdot10^{-24\pm 1}$ cm$^2$ GeV$^{-1}$ which gives 
(for $m_A\sim 10^{-3}$ eV) $\sigma\sim 10^{-36\pm 1}$ cm$^2$ 
i.e in the region of cross sections mediated by the exchange of heavy neutrinos.
In conclusion, and subject to our assumptions on quantum gravity induced mass 
generation for the minoron we have shown that the CSM can offer a viable scenario
for the explanation of Dark Matter.

\section{Outlook}
To conclude we summarize the main features of the CSM elaborated in this paper:

\bit
\item There is a range of parameter values for which the CSM is perturbative and
         the electroweak vacuum remains stable for all energies up to $\MPL$.
\item The main prediction of the model is a new and almost sterile scalar resonance
         which comes with low mass heavy neutrinos, but nothing else. 
\item All new degrees of freedom are very weakly coupled to SM matter.
\item There exist Casa-Ibarra matrices $\RCI$
         for which resonant leptogenesis is possible. 
\item The pseudo-Goldstone boson associated with the breaking of $(B-L)$ (`minoron')
         is a possible Dark Matter candidate, whose non-vanishing mass is an indirect
         manifestation of quantum gravity. 
\eit

We stress again that these properties set the CSM apart from many other current 
proposals (such as SUSY Higgs, two doublet models, vector-like models) where neutral 
scalars are usually accompanied by other and `non-sterile' charged excitations, and which 
would all have to be produced together. So, barring the inconvenient possibility that a 
new scalar could escape detection because the associated resonance could 
be too narrow for the LHC energy bins, the acid test of the present model 
will be whether or not the new scalar  shows up in future LHC searches with 
increased luminosity. In this way the model is eminently falsifiable.

\vspace{0.5cm}

\noindent{\bf {Acknowledgments:}}  We thank P. Chankowski for discussions. 
K.A.M. thanks the AEI for hospitality and partial support during this work. K.A.M. was partially supported by the Polish National Science Center grant
DEC-2017/25/B/ST2/00165.

\vspace{3mm}

\section*{Appendix: Field Operators for Massive and Massless Neutrinos}

Since the use of the Weyl fields in $S$-matrix calculations is perhaps not so common, 
for readers' convenience we here give the explicit decompositions of the corresponding 
field operators in terms of creation and annihilation operators (see {\em e.g.}
\cite{BaggerWess} for an introduction to SL(2,$\mathbb{C}$) spinor formalism).
These expressions
can be derived for instance following Weinberg's procedure \cite{WMassive,WMassless} 
(although we use different normalization conventions), and they are the ones used in 
the computation of the matrix elements $\langle h\nu|\cL_Y|N\rangle$ and
$\langle h\bar\nu |\cL_Y|N\rangle$ required for the determination of the 
CP asymmetries in (\ref{Eq:ep-def-nowa}).

For the massive case, and suppressing family indices
the field operator $\bar{N}^{\da}(x)$ in the fundamental representation takes the form
\eqs{\nn
\bar{N}^{\da}(x)
&=&
\sum_{r=\pm 1/2}  \int\!\frac{{\rm d}^3{\bf{p}}}{2(2\pi)^3 \sqrt{m^2+{\bf p}^2}} \times \nn\\[2mm]
&& \!\!\!\!\!\!\!\!\!\!\!\!\!\!\!\!\!\!\!\!\!\!\!\!
\times\, \Big\{
\bar{u}^{\da}_r({\bf{p}})  b_r({\bf{p}}) e^{-i p_\mu x^\mu} 
+\;   \bar{v}^{\da}_r ({\bf{p}})  b_r^\dagger ({\bf{p}}) e^{i p_\mu x^\mu}  \Big\} \; ,
\nn
}
where $p^\mu = (p^0, {\bf p})$ is on shell: $p^0=\sqrt{m^2+{\bf p}^2}$. The normalization
of creation and annihilation operators can be read off from the anti-commutator
\eq{\nn
\big[ b_r({\bf{p}}),\ b_s^\dagger ({\bf{q}}) \big]_{+}
=
2\sqrt{m^2+{\bf{p}}^2}\ \delta_{rs}(2\pi)^3 \delta^{(3)}({\bf{p}}-{\bf{q}}),
}
The two-component spinor wave functions $\bar{u}^{\da}_r ({\bf{p}})$ and $\bar{v}^{\da}_r({\bf{p}})$ 
can be likewise read off as the column vectors 
\eq{
{\bar{u}}({\bf p}) \,= \, B_{\bf{p}}, \qquad
{\bar{v}}({\bf p})\,=\, \, B_{\bf{p}}\,\epsilon^{-1}\,.
}
from the $2\times 2$ matrix 
\eq{\label{Eq:B_p-massive}
B_{\bf{p}} \,\equiv \,\frac{1}{\sqrt{2(m+p^0})} [p^\mu \anti{\sigma}_\mu+m\mathds{1}].
}
where $\epsilon^{-1}$ is the inverse antisymmetric metric, and where $r,s=\pm 1/2$ label the eigenvalues of 
$J_3$ in the rest frame. The conjugate Weyl spinor operator is obtained by taking 
the hermitean conjugate $N_\al \equiv \epsilon_{\al\be} \bar{N}^{\db\dagger}$.

These formulae are, of course, in complete accord with textbook formulas in 4-spinor notation.
More precisely, combining $N_\al$ and $\bar{N}^\da$ into a Majorana spinor as in (\ref{Eq:Majorana})
we reproduce the standard formula
\eqs{\nn
\psi_{N}(x)
&=&
\sum_{r=\pm{1/2}}  \int\!\frac{{\rm d}^3{\bf{p}}}{2(2\pi)^3 \sqrt{m^2+{\bf p}^2}} \times \nn\\[2mm]
&& \!\!\!\!\!\!\!\!\!\!\!\!\!\!\!\!\!\!\!\!\!\!\!\!
\times  \Big\{ {U}^{}_r ({\bf{p}}) b_r ({\bf{p}}) e^{-i p_\mu x^\mu} +\,
  {V}^{}_r({\bf{p}})  b_r^\dagger ({\bf{p}}) e^{i p_\mu x^\mu}
\Big\},
}
with the 4-spinors $U\equiv (v_\al, \bar{u}^\da)$ and 
$V \equiv (u_\al, \bar{v}^\da)$ obeying the completeness relations
\eqs{\label{Eq:U-V-norm}
\sum_{r} U_r({\bf p})\anti{U}_r({\bf p})&=&\ds{p}+m\nn\\
\sum_{r} V_r({\bf p})\anti{V}_r({\bf p})&=&\ds{p}-m\,,
}
with the Weyl representation of Dirac matrices, see \cite{BaggerWess}. 
Although with this normalization the limit $m\to0$ is non-singular, a more physical
choice of basis for massless spinors corresponds to the helicity eigenstates, rather 
than just the formal limit of the above expressions. In this basis we have for massless spinors
\eqs{\nn
\anti{\nu}^{\da}(x)
&=&
\!\int\!\frac{{\rm d}^3{\bf{p}}}{2(2\pi)^3 |{\bf p}|}\,
\bar{w}^{\da}({\bf{p}}) \times  
\nn\\[2mm]
&&
\times\, \Big\{
a_+ ({\bf{p}}) e^{-i p_\mu x^\mu} - \, a_-^\dagger ({\bf{p}}) e^{i p_\mu x^\mu} \Big\}\,.
\nn}
Because of the degeneracy of the Weyl operator $p_\mu \si^\mu$ in the massless case
there is now {\em only one} spinor wave function, unlike for the massive case where 
there are two. This helicity wave function satisfies the Weyl equation 
$p_\mu \si^\mu_{\al\db} \bar{w}^\db ({\bf{p}}) = 0$ and obeys the completeness relation
\eq{
\bar{w}^{\da}({\bf{p}})\, {w}^{\be}({\bf{p}}) = p_\mu \anti\si^{\mu\,{\da\be}}\,.
}
The helicity eigenstates are
\beq
|\nu({\bf{p}})\rangle = a_-^\dagger ({\bf{p}}) |0\rangle    \;,\quad
|\bar\nu({\bf{p}})\rangle = a_+^\dagger ({\bf{p}}) |0\rangle
\eeq
whence $a_-^\dagger ({\bf{p}})$ creates a helicity --1/2 neutrino, while  
$a_+^\dagger ({\bf{p}})$ creates a helicity +1/2 antineutrino. 
Notice also, that the associated 4-spinors appearing in the resulting 
decomposition of the Majorana field $\psi_\nu$ in (\ref{Eq:Majorana})
are consistent with the massless limit of \refer{Eq:U-V-norm}.



\end{document}